\documentclass[aps,preprint]{revtex4-1}

\usepackage{amssymb}
\usepackage{amsmath}
\usepackage{graphicx}
\usepackage{dcolumn}
\usepackage{bm}
\usepackage{mathtools}

\usepackage[ruled]{algorithm2e}
\usepackage{color}

\DeclarePairedDelimiter\bra{\langle}{\rvert}

\DeclarePairedDelimiter\ket{\lvert}{\rangle}

\DeclarePairedDelimiterX\braket[2]{\langle}{\rangle}{#1 \delimsize\vert #2}

\newcommand{\vB}{{\vec{B}}}

\newcommand{\vE}{{\vec{E}}}

\newcommand{\vecr}{{\vec{r}}}

\newcommand{\vJ}{{\vec{J}}}

\newcommand{\vnabla}{{\vec{\nabla}}}

\newcommand{\hrho}{{\hat{\rho}}}

\newcommand{\hH}{{\hat{H}}}

\newcommand{\hA}{{\hat{A}}}

\newcommand{\hP}{{\hat{\mathcal{P}}}}

\draft 

\begin{document}

	
	
	

	\title{Mixed Quantum-Classical Electrodynamics: Understanding  Spontaneous Decay and Zero Point Energy} 

	
	
	
	

	

	\author{Tao E. Li}
	\affiliation{Department of Chemistry, University of Pennsylvania, Philadelphia, Pennsylvania 19104, USA}
	
	
	

	\author{Abraham Nitzan}
	\affiliation{Department of Chemistry, University of Pennsylvania, Philadelphia, Pennsylvania 19104, USA}
	\author{Maxim Sukharev}
	\affiliation{Department of Physics, Arizona State University, Tempe, Arizona 85287, USA}
	\affiliation{College of Integrative Sciences and Arts, Arizona State University, Mesa, AZ 85212, USA}
	\author{Todd Martinez}
	\affiliation{Department of Chemistry and The PULSE Institute, Stanford University, Stanford, California 94305, USA}
	\affiliation{SLAC National Accelerator Laboratory, 2575 Sand Hill Road, Menlo Park, CA 94025, USA}
	\author{Hsing-Ta Chen}
	\affiliation{Department of Chemistry, University of Pennsylvania, Philadelphia, Pennsylvania 19104, USA}
	\author{Joseph E. Subotnik}
	\affiliation{Department of Chemistry, University of Pennsylvania, Philadelphia, Pennsylvania 19104, USA}

	
	
	

	\date{\today}

	\begin{abstract}
		
		The dynamics of an electronic two-level system coupled to an electromagnetic field are simulated explicitly for one and three dimensional systems through semiclassical propagation of the Maxwell-Liouville equations.  We consider three flavors of mixed quantum-classical dynamics: the classical path approximation (CPA), Ehrenfest dynamics, and symmetrical quantum-classical (SQC) dynamics.  The CPA  fails to recover a consistent description of spontaneous emission.  A consistent ``spontaneous'' emission can be obtained from Ehrenfest dynamics--provided that one starts in an electronic superposition state.  Spontaneous emission is always obtained using SQC dynamics.  Using the SQC and Ehrenfest frameworks, we further calculate the dynamics following an incoming pulse, but here we find very different responses: SQC and Ehrenfest dynamics deviate sometimes strongly in the calculated rate of decay of the transient excited state. Nevertheless, our work confirms the earlier observations by W. Miller [\textit{J. Chem. Phys.} \textbf{69}, 2188-2195, 1978] that Ehrenfest dynamics can effectively describe some aspects of spontaneous emission and highlights new possibilities for studying light-matter interactions with semiclassical mechanics.
		
	\end{abstract}

	
	\maketitle 

	\section{Introduction}
	
	Understanding the dynamics of light-matter interactions is essential for just about any flavor of physical chemistry; after all, with a few exceptions, photons are the most common means nowadays to interrogate molecules and materials in the laboratory.  Today, it is standard to  study molecules and materials with light scattering experiments, absorption spectroscopy, pump-probe spectroscopy, etc. For a chemist, the focus is usually on the matter side, rather than the electromagnetic (EM) field side: one usually pictures an incoming EM field as a time-dependent perturbation for the molecule. Thereafter, one calculates how the molecule responds to the perturbation and, using physical arguments and/or semiclassical insight, one extrapolates how the molecular process will affect the EM field.  For instance, in an absorption experiment, we usually assume linear response theory\cite{kamakoffbookonline} when calculating how much energy the molecule absorbs. More precisely, one calculates a dipole-dipole correlation function and then, after Fourier transform, one can make an excellent prediction for the absorption pattern. For weak electric fields, this approach often results in reliable data.
	
	However, in many situations involving strong light/matter interactions (e.g. laser physics), the states of the radiation field and the material sub-systems have to be considered on equal footing. An example of strong recent interest is the host of observed phenomena that manifest strong exciton-photon coupling.\cite{torma2014strong, sukharev2017optics, vasa2017strong} Closely related, and also in recent focus, are observations and models pertaining to strong interactions between molecules and electromagnetic modes confined in optical cavities.\cite{khitrova2006vacuum, gibbs2011exciton, lodahl2015interfacing} As another example,  recent studies by Mukamel\cite{mukamel:2017:comment_prl}, Bucksbaum\cite{bucksbaum:2016:prl} and coworkers who have explored the proper interpretation of x-ray pump-probe scattering experiments and, in particular, the entanglement between electrons, nuclei and photons. Beyond the analysis of simplified quantum models, the important tools in analyzing many of these phenomena  are variants of coupled Maxwell and Schr\"odinger (or, when needed, quantum-Liouville) equations, where the radiation field is described by  classical Maxwell equations while the molecular system is modeled with a handful of states and described quantum mechanically.\cite{sukharev2017optics, Csesznegi1997, Su2011, Zhang2012, Puthumpally-Joseph2015, Sukharev2011, Smith2017, masiello2005dynamics, lopata2009nonlinear}. A classical description of the radiation field is obviously an important element of simplification in this approach, which makes it possible to simulate the optical response of realistic model systems. However, open questions remain in this area, in particular:
	\begin{itemize}
		\item How does spontaneous emission emerge, if at all, in semiclassical calculations?
        \item How do we best describe computationally the possibly simultaneous occurrence of absorption, scattering, fluorescence and non-linear optical response following a pulse or CW excitation of a given molecular system that may interact with its environment?
         \item How do we treat both quantum-mechanical electron-electron interactions (e.g. spin-orbit coupling) and classical electronic processes (e.g. electronic energy transfer)  in a consistent fashion?
	\end{itemize}

	In the future, our intention is to address each and every one of these questions.  For the present article, however, our goal is to address the first question.
	We note that spontaneous emission rates can be evaluated from the rate of energy emission by a classical dipolar antenna\cite{gersten2005topics}. An important quantification of this observation has been provided by Miller\cite{Miller1978} who has shown that apart from semiclassical corrections, spontaneous decay rates can be ascertained from classical dynamics. Indeed, for a dipolar harmonic oscillator Miller has shown that a semiclassical decay rate can be obtained from classical dynamics exactly. His treatment\cite{Miller1978}, however, raises several questions.
	First, in Ref. \onlinecite{Miller1978}, the molecular system is represented by a classical harmonic oscillator rather than a 2-level system.  How will the observations made by Miller be affected with a proper quantum-mechanical treatment? What will  be the performance of  mixed semiclassical treatments for spontaneous emission, and which semiclassical treatment will perform best?  Second, in Ref. \onlinecite{Miller1978}, no explicit light pulses are applied to the  electronic system, but one can ask: If a pulse of light is applied to the system, and we use mixed quantum-classical dynamics, is the propagated photon field consistent with the ensuing molecular dynamics? With an external temperature, do we recover detailed balance? In this article, we will address most of these questions, paying special attention to the recent symmetrical quantum-classical (SQC) dynamics protocol of Cotton and Miller\cite{cotton2013symmetrical, Cotton2013}.
	
	This article is arranged as follows. In Sec. \ref{sec:spontaneous_decay}, we briefly review the theory of spontaneous decay. In Sec. \ref{sec:semiH}, we introduce the semiclassical Hamiltonian in our model. In Sec. \ref{sec:methods}, we implement  Ehrenfest dynamics, CPA and SQC. In Sec. \ref{sec:simulation_details}, simulation details are given. In Sec. \ref{sec:result}, we compare  results for spontaneous decay. In Sec. \ref{sec:discussion}, we simulate and analyze two cases: $(i)$ the arrival of an incoming pulse and $(ii)$ dephasing effects. We conclude in Sec. \ref{sec:conclusion}.
	
	For notation, we use the following conventions: $\hbar \omega_0$ is used to represent the energy difference between the excited state $\ket{e}$ and the ground state $\ket{g}$; $\hbar \omega_{k'}$ (or $\hbar c k'$) is used to represent the energy of the photon with wave vector $k'$; $\mu_{12}$ is the electric transition dipole moment of the molecule;  $\sqrt{\frac{1}{a}}$ represents the molecular size so that the transition dipole moment with a characteristic charge $q$ is approximately $\mu_{12}\approx\sqrt{\frac{1}{a}}q$;
	 $\varsigma$ is used to represent a dephasing rate; $U_0$ denotes the total energy of an incident pulse; $k_0$ denotes the peak position, in Fourier space, of an incident pulse; $b$ is a parameter fixing the width of an incident pulse in space; and $c$ is the speed of light. We work below in SI units.

	\section{Theory of Spontaneous Emission}\label{sec:spontaneous_decay}
	
	For completeness, and because we will work in both one and three dimensions, it will be convenient to briefly review the theory of spontaneous emission and dipole radiation. Consider a molecular species in an excited state $\ket{e}$ which can decay to the ground state $\ket{g}$  by emitting a photon spontaneously.  
	
	\subsection{The Fermi's Golden Rule (FGR) Rate}
	Let the vacuum state for the radiation field be $\ket{0}$. Suppose that  initially the system is in state $\ket{e}\otimes\ket{0}$.  At long times, we expect to observe spontaneous emission, so that the final state will be 	$\ket{g}\otimes a^{\dagger}_{q, s}\ket{0}$.   Here,  $a^{\dagger}_{q, s}$ creates a photon with wave vector $\vec{q}$ and polarization $s$.

	We now apply Fermi's Golden Rule (FGR) for the emission rate.  We further make the dipole approximation, so that the interaction Hamiltonian for a molecule sitting at the origin  is $H_{int} = -q \hat{\vec{r}} \cdot \hat{\vec{E}}(0)$, where $q $ is the electronic charge, $\hat{\vec{r}}$ is the position operator for the quantum system, and $\hat{\vec{E}}(0)$ is the electric field at the origin.
	In such a case, the decay rate $k$ in 3D can be calculated as follows\cite{Schwabl2007}:
	\begin{subequations}
	\begin{align}
	k_{\text{3D}} &= \frac{2\pi}{\hbar^2} \sum_{\vec{k}', \vec{s}} \frac{\hbar \omega_{k'}}{2 \epsilon_0 V} |\vec{\mu}_{12}\cdot \vec{\epsilon}_{\vec{k}', \vec{s}}|^2 \delta(\omega_0 - ck') \label{eq:FGR_3d_before} \\
	&  = \frac{2\pi}{\hbar^2} 2\int d\varphi\int \sin\theta d\theta \int k'^2 dk'\frac{V}{(2\pi)^3} \times \nonumber \\
	& \ \ \ \frac{\hbar \omega_{k'}}{2\epsilon_0 V} |\mu_{12}|^2\cos^2\theta \delta(\omega_0 - ck')\label{eq:FGR_3d_before2} \\
	&= \frac{\omega_0^3 |\mu_{12}|^2}{3\pi \epsilon_0 c^3\hbar}\label{eq:FGR_3d}
	\end{align}
	\end{subequations}
	Here, $\mu_{12} = |\bra{e}q \hat{\vec{r}} \ket{g}|$ is the three-dimensional transition dipole moment of the molecule, $\vec{\epsilon}_{\vec{k}', \vec{s}}$ is the a unit vector in direction of the electric field indexed by the wave vector $\vec{k}'$ and the polarization vector $\vec{s}$, and $\hbar \omega_0$ is the energy difference between $\ket{e}$ and $\ket{g}$.  Eqn. (\ref{eq:FGR_3d_before}) is the usual FGR expression. In Eqn. (\ref{eq:FGR_3d_before2}), if we replace the discrete $\sum_{\vec{k}'}$   with the continuous $\int d\varphi  d\theta dk' \sin\theta k'^2 \rho(k')$, where $\rho(k') = V/(2\pi)^3$ is the three-dimensional density of states (DOS) for the photons, we recover Eqn. (\ref{eq:FGR_3d}).
	
	In what follows below, it is useful to study EM radiation in 1D as well as in 3D. To that end, we will imagine charge distributions that are function of $x$ only, i.e. they are uniform in $y$ and $z$ directions. In 1D, the density of states (DOS) for the photon field is $\rho(\vec{k'}) = L_x/2\pi$. Therefore, the decay rate in 1D is:
	\begin{subequations}\label{eq:FGR_1d_all}
	\begin{align}
	k_{\text{1D}}  &= \frac{2\pi}{\hbar^2} \sum_{k_x', s} \frac{\hbar \omega_{k_x'}}{2 \epsilon_0 V} |\mu_{12}|^2 \delta(\omega_0 - ck_x')  \\
	&  = \frac{2\pi}{\hbar^2} 2\int dk_x'\frac{L_x}{2\pi} \frac{\hbar \omega_{k_x'}}{2\epsilon_0 V} |\mu_{12}|^2 \delta(\omega_0 - ck_x')
	\end{align}
	\end{subequations}
	Using $V = L_xL_yL_z$ and defining the one-dimensional dipole moment $|\mu_{12}^{1D}|^2 = |\mu_{12}^{3D}|^2/L_yL_z$, we can rewrite the final 1D rate as
	\begin{eqnarray}
		k_{\text{1D}} = \frac{\omega_0}{ \hbar \epsilon_0 c } |\mu_{12}^{1D}|^2  \label{eq:FGR_1d}
	\end{eqnarray}
	Below,  we will use $\mu_{12}$ to represent either $\mu_{12}^{1D}$ and $\mu_{12}^{3D}$ depending on context.

	Note that, in 1D, the spontaneous decay rate $k_{\text{1D}}$ depends linearly on the frequency $\omega_0$ and quadratically on the transition dipole moment $\mu_{12}$. In 3D, however, $k_{\text{3D}}$ depends cubically on $\omega_0$ instead of linearly, but still quadratically  on $\mu_{12}$.  Note that, for Eqns. (\ref{eq:FGR_3d}) and (\ref{eq:FGR_1d}) to apply, two conditions are required: $(i)$ The dipole approximation must be valid, i.e. the wavelength of the spontaneous light must be much larger than width of molecule.  $(ii)$ The coupling between molecule and radiation field must be weak to ignore any feedback of the EM field, i.e. $\omega_0$ must be much larger than the inverse lifetime.
	
	\subsection{The Abraham-Lorentz Rate}\label{sec:AL}
	While FGR is the standard protocol for modeling spontaneous emission with quantum mechanics, we can also recover  a similar decay rate with classical mechanics by using the Abraham-Lorentz equation\cite{daboul1974generalising} . For a classical charged harmonic oscillator moving in the $x$ direction with mass $m$, the Abraham-Lorentz equation reads
	\begin{eqnarray}\label{eq:ODE_Abraham_Lorentz}
	m\ddot{\vec{x}}(t) = -m\omega_0^2 \vec{x}(t) + m\tau \dddot{\vec{x}}(t)
	\end{eqnarray}
	where $\tau = q^2/6\pi\epsilon_0 c^3 m$ has the dimension of time. The last term in Eqn. (\ref{eq:ODE_Abraham_Lorentz}) represents the recoil force on a particle as it feels its own self-emitted EM field. Since $\tau \ll 1/\omega_0$, we can assume the damping effect is small and so we replace $m\tau \dddot{\vec{x}}(t)$ by $-m\omega_0^2\tau \dot{\vec{x}}(t)$ to obtain
	\begin{eqnarray}\label{eq:ODE_Abraham_Lorentz2}
	m\ddot{\vec{x}}(t) = -m\omega_0^2 \vec{x}(t) - m\omega_0^2\tau \dot{\vec{x}}(t)
	\end{eqnarray}
	Eqn. (\ref{eq:ODE_Abraham_Lorentz2}) represents a damped harmonic oscillator, which has a well-know solution
	\begin{eqnarray}\label{eq:AL_solution}
	\begin{split}
	\vec{x}(t) &= x_0\cos\left(\omega_0\sqrt{1- \frac{\omega_0^2\tau^2}{4}} t + \phi\right)e^{-\frac{k_{\text{AL}}}{2}t}\hat{e}_x \\
	&\approx x_0\cos\left(\omega_0 t + \phi\right)e^{-\frac{k_{\text{AL}}}{2}t}\hat{e}_x
	\end{split}
	\end{eqnarray}
	since $\omega_0 \tau \ll 1$. In Eqn. (\ref{eq:AL_solution}), the amplitude $x_0$ and the phase $\phi$ will depend on the initial conditions, and the decay rate $k_{\text{AL}}$  is
	\begin{eqnarray}
	\begin{split}
	k_{\text{AL}} = \omega_0^2\tau = \frac{q^2\omega_0^2}{6\pi\epsilon_0 c^3 m}
	\end{split}
	\end{eqnarray}
	At this point, we can write down the total energy of the harmonic oscillator:
	\begin{eqnarray}
	\begin{split}
	 \mathcal{E}(t) &= \frac{1}{2}m\omega_0^2  \vec{x}^2(t) + \frac{1}{2}m\dot{\vec{x}}^2(t) \\
		&= m\omega_0^2 x_0^2e^{-k_{\text{AL}}t} \left( 1 + \frac{1}{8}\frac{k_{\text{AL}}^2}{\omega_0^2}\cos^2\left(\omega_0 t + \phi\right)\right)\\
		&\approx m\omega_0^2 x_0^2 e^{-k_{\text{AL}}t}
	\end{split}
	\end{eqnarray}
	To relate the Abraham-Lorentz rate $k_{\text{AL}}$ to the FGR rate in 3D, we require a means to connect a classical system with mass $m$ to a pair of quantum mechanical states. To do so, we imagine the oscillator is quantized and that the motion is occurring in the ground state, where $\sqrt{x_0^2} =  \sqrt{\hbar/2m\omega_0}$. This is equivalent to asserting that the initial energy of the dipole is $\frac{1}{2}\hbar\omega_0$, which we set equal to the total dipole energy, $m \omega_0^2  x_0^2 $.
	If we further assert that the dipole operator is off-diagonal (as in Eqn. (\ref{eq:Pr})), we may substitute $qx_0 \approx \mu_{12}$, which leads to the following Abraham-Lorentz rate ($k_{\text{AL}}$)
	\begin{eqnarray}\label{eq:k_AL}
	k_{\text{AL}} =\frac{q^2 x_0^2 \omega_0^3}{3\pi\epsilon_0 c^3 \hbar} = \frac{|\mu_{12}|^2 \omega_0^3}{3\pi\epsilon_0 c^3 \hbar} = k_{\text{FGR}}^{\text{3D}} 
	\end{eqnarray}
	With this ansatz, the Abraham-Lorentz decay rate $k_{\text{AL}}$ is equal to the FGR rate in 3D. Note that several \textit{ad hoc} semiclassical assignments must be made for this comparison, and it is not clear how to generalize the Abraham-Lorentz approach to treat more than two electronic states in a consistent fashion. 
	
	\subsection{The Asymptotic Electromagnetic Field}
	Below, we will analyze different schemes for solving Maxwell's equations coupled together with the Liouville equation, and it will be helpful to compare our results with the standard theory of dipole radiation.
	According to classical electrodynamics, if a dipole is located at the origin and is driven by an oscillating field, the electromagnetic (EM) field is generated with the energy density  (at time $t$ and position $\vecr$) given in the far-field by\cite{Griffiths2012}
	\begin{equation}\label{eq:dipole_radiation_1}
	u(\vecr, t) = \frac{\mu_0}{c^2}\frac{\omega_0^4\mu_{12}^2}{16\pi^2}\frac{\sin^2\theta}{r^2}\cos^2(\omega_0(t-r/c)).
	\end{equation}
	Here, without loss of generality, we assume that the dipole is pointing in the $z-$direction, so that $\theta$ is the polar angle from the $z$-axis. $r$ is the distance from the observer to the dipole (sitting at the origin). 
 	Eqn. (\ref{eq:dipole_radiation_1}) predicts that, for the energy density, there is $\sin^2\theta$ dependence on the polar angle $\theta$ and $1/r^2$ dependence on the distance $r$. Note that Eqn. (\ref{eq:dipole_radiation_1}) is valid in the far-field when $r \gg \lambda \gg d$, where $\lambda$ is the wavelength of EM field and $d$ is the size of the dipole. 

	\section{The Semi-classical Hamiltonian}\label{sec:semiH}
	We consider the problem of a two-level system coupled to a radiation field. After a  Power-Zienau-Woolley transformation\cite{mukamel1999principles, cohen1997photons} is applied, the Hamiltonian reads as follows:
	\begin{eqnarray}\label{eq:original_H}
	\begin{split}
	\hH = &\hH_s + \frac{1}{2}\int  d\vecr \left[ \frac{1}{\epsilon_0}\hat{D}^{\perp}(\vecr)^2 + \frac{1}{\mu_0}\hat{B}(\vecr)^2 
	 \right]\\
				& - \int d\vecr \frac{\hat{D}^{\perp}(\vecr)}{\epsilon_0}\cdot \hP^{\perp}(\vecr) + \frac{1}{2\epsilon_0}\int d\vecr |\hP^{\perp}(\vecr)|^2 
	\end{split}
	\end{eqnarray}
	Here, $\hat{B} = \nabla \times \hA$, $\hat{D}^{\perp} = \epsilon_0\hat{E} + \hP^{\perp} $. $\hA$ is the vector potential for the  EM field and $\hP^{\perp} $ is the polarization operator for the  matter.
	For the EM field, the relevant commutators are: $[\hat{D}^{\perp}(\vecr), \hat{A}(\vecr')] =i\hbar \delta^{\perp}(\vecr - \vecr')$, where  $\delta^{\perp}$ is the transverse delta function. $H_s$ is the Hamiltonian of the electronic system, which will be defined below. We ignore all magnetic moments in Eqn. (\ref{eq:original_H}).

	Eqn. (\ref{eq:original_H}) is a large Hamiltonian, written in the context of a quantum field. For semiclassical dynamics, it is convenient to extract the so-called ``electronic Hamiltonian'' that depends only parametrically on the EM field. Following Mukamel\cite{mukamel1999principles}, one route to achieve such a semiclassical Hamiltonian is to consider the equation of motion for an observable of the matter $\hat{Q}$:

	\begin{equation}
	\begin{aligned}
	\frac{\hbar}{i}\frac{d\hat{Q}}{dt} &= \left[ \hH_s, \hat{Q}\right] - \frac{1}{2}\int d\vecr  \left( \left[\hP^{\perp}, \hat{Q} \right]\hat{E}^{\perp} + \hat{E}^{\perp}\left[\hP^{\perp}, \hat{Q} \right]        \right)  \\
	&\stackrel{?}{=} \left[ \hH^{el}, \hat{Q}\right]
	\end{aligned}
	\end{equation}

	If we  approximate that the E-field is classical, so that we may commute $\hat{E}^{\perp}$ with all matter operators, we find the following semiclassical electronic Hamiltonian:
	\begin{eqnarray}\label{eq:semi-H}
	\hH^{el}(E) = \hH_s - \int d\vecr\  \vE^{\perp}(\vecr)\cdot \hP^{\perp}(\vecr)
	\end{eqnarray}
	With only one charge center, however, we will not need to distinguish between the longitudinal and perpendicular components, and so we will drop the $^{\perp}$ notation below.
	
	For this paper, we consider the simplest case of two electronic states: the ground state  $\ket{g}$ and the electronic excited state $\ket{e}$.  Thus, we  represent $H_s$ as follows:
	\begin{equation}\label{eq:Hs}
	\hH_s =
	\begin{pmatrix}
	0 & 0\\ 0 & \hbar \omega_0 
	\end{pmatrix}
	\end{equation}
	Furthermore, we assume that ($a$) these states carry no permanent dipole and ($b$) the transition between them is characterized by two single electron orbitals $\psi_g$ and $\psi_e$ and an effective charge $q$ such that the transition dipole density is given by
	\begin{equation}\label{xi}
	\vec{\xi}(\vecr) = q\vecr \cdot \psi_e^{\ast}(\vecr)\psi_g(\vecr)
	\end{equation}
	with a corresponding polarization operator:
	\begin{equation}\label{eq:Pr}
	\hP(\vecr) =
	\begin{pmatrix}
	0 & 1\\ 1 & 0 
	\end{pmatrix}
	\vec{\xi}(\vecr)
	\end{equation}
	
	For example, in 3D,  in the common case that $\psi_e(\vecr)$ is a  $p_z$ orbital ($\frac{2^{1/2}a^{5/4}}{\pi^{3/4}}ze^{-ar^2/2}$)  and $\psi_g(\vecr)$ is an $s$ orbital ($(\frac{a}{\pi})^{3/4}e^{-ar^2/2}$), $\vec{\xi}(\vecr)$ would be
	\begin{equation}\label{eq:3dxi_0}
	\vec{\xi}^{3D}(\vecr) = q\vecr \frac{2^{1/2} a^{2}}{\pi^{3/2} } ze^{-ar^2}
	\end{equation}
	If we consider a charge distribution that is effectively 1D, changing along in the $x$ direction but polarized in the $z$ direction, the reduced form of $\vec{\xi}(\vecr)$  would be
	\begin{equation}\label{eq:1dxi_0}
	\vec{\xi}^{1D}(x) =q \hat{e}_z \frac{1}{\sqrt{2\pi}}  e^{-ax^2}
	\end{equation}
	
	The magnitude of $\vec{\xi}(\vecr)$ is related to the magnitude of the total transition dipole moment, $\vec{\mu}_{12}$:
	\begin{equation}\label{eq:mur}
	\mu_{12} =|\vec{\mu}_{12}| = |\bra{e}q\hat{\vec{r}} \ket{g}| = |\int d\vecr\  \vec{\xi}(\vecr)|
	\end{equation}
	Eqn. (\ref{eq:mur}) guarantees that, when the width of $\hP(\vecr)$ approaches $0$, Eqn. (\ref{eq:semi-H}) becomes the standard
	dipole Hamiltonian, $\hH^{el} = \hH_s - \vec{\mu}_{12}\cdot\vE(0)$. This definition allows us to rewrite Eqns. (\ref{eq:3dxi_0}-\ref{eq:1dxi_0}) above, as follows:
	\begin{subequations}
	\begin{align}
	\vec{\xi}^{3D}(\vecr) &= \frac{2 a^{5/2}}{\pi^{3/2} }\mu_{12} \vecr ze^{-ar^2}\label{eq:3dxi}\\
	\vec{\xi}^{1D}(x) &=\sqrt{\frac{a}{\pi}} \mu_{12}\hat{e}_z e^{-ax^2}\label{eq:1dxi}
	\end{align}
	\end{subequations}
	Note that $\vec{\xi}^{3D}$ and $\vec{\xi}^{1D}$ have different units. 
	
	In Appendix A we will show that under the point dipole limit -- where the width of $\vec{\xi}(\vecr)$ is much smaller than the wavelength of EM field, so that  $\vec{\xi}(\vecr)$ can be treated as a delta function -- some analytic results can be derived for the coupled electronic-photons dynamics.

	\section{Methods}\label{sec:methods}

	Many mixed quantum-classical semiclassical dynamics tools have been proposed over the years to address coupled nuclear-electronic dynamics, including wave packet dynamics\cite{lee1982exact, cina2003wave}, Ehrenfest dynamics\cite{Li2005}, surface-hopping dynamics\cite{Tully1990, Nielsen2000}, multiple spawning dynamics\cite{Ben-Nun2000}, and partially linearized density matrix dynamics (PLDM)\cite{Huo2012}. Except for the Ehrenfest (mean-field) dynamics, other methods are usually based on the Born-Oppenheimer approximation, which relies on the timescale separation between (slow) classical and (fast) quantum  motions. Such methods cannot be applied in the present context because the molecular timescales and the relevant photon periods are comparable.\footnote{There is one interesting nuance in this argument. The standard approach for embedding a quantum DOF in a classical environment is the quantum classical Liouville equation(QCLE), which can be approximated by PLDM\cite{Huo2012} or surface-hopping dynamics\cite{Tully1990}. In the present case, for photons interacting with a handful of electronic states, the Hamiltonian is effectively a spin-boson Hamiltonian, which is treated exactly by the QCLE, regardless of the Born-Oppenheimer approximation or any argument about time-scale separation. Nevertheless, in general, we believe that many semi-classical dynamics, especially surface-hopping dynamics, will not be applicable in the present context.} The Ehrenfest approximation relies on the absence of strong correlations between interacting subsystems, and may be valid under more lenient conditions. We therefore limit the following discussion to the application of the Ehrenfest approximation and its variants\footnote{Note that in most applications the Ehrenfest approximation is used to describe coupled electronic and nuclear motions where timescale separation determines the nature of the ensuing dynamics. Here we use this approximation in the spirit of a time dependent Hartree (self consistent field) approximation. Since timescale separation is not invoked, the success of this approach should be scrutinized by its ability to describe physical results, as is done in the present work.
	}.
	



	\subsection{Ehrenfest Dynamics}

	According to Ehrenfest dynamics for a classical radiation field and a quantum molecule, the molecular density operator $\hrho(t)$ is propagated according to
	\begin{equation}\label{density_matrix_evolution}
	\frac{d}{dt}\hrho(t) = -\frac{i}{\hbar}[\hH_s - \int d\vecr\  \vE(\vecr, t)\cdot \hP(\vecr),\  \hrho(t) ]
	\end{equation}
	while the time evolution of the radiation field is given by the Maxwell's equations
	\begin{eqnarray}\label{Maxwell}
	\begin{split}
	\frac{\partial \vB(\vecr)}{\partial t} &= -\vnabla \times \vE(\vecr) \\
	\label{equ:EM2}
	\frac{\partial \vE(\vecr)}{\partial t} &= c^2\vnabla \times \vB(\vecr) - \frac{\vJ(\vecr)}{\varepsilon_0} 
	\end{split}
	\end{eqnarray}
	Here, the current density operator, $\hat{\vJ} = d\hat{P}/dt$, is replaced by its expectation value:
	\begin{equation}\label{current_J}
	\vJ(\vecr) = \frac{d}{dt} \text{Tr}(\hrho \hP(\vecr))
	\end{equation} 
	If we substitute Eqns. (\ref{eq:Pr}) and (\ref{density_matrix_evolution}) into Eqn. (\ref{current_J}), the current density $\vJ(\vecr)$ can be simplified to 
	\begin{equation}\label{eq:J_simplified}
	\vJ(\vecr) = -2\omega_0\text{Im}(\rho_{12})\vec{\xi}(\vecr)
	\end{equation}
	where $\rho_{12}$ is the coherence of  the density matrix $\rho$.
	
	Two points are noteworthy: First, because Eqn. (\ref{density_matrix_evolution}) does not include any dephasing or decoherence, there is also an equivalent equation of motion for the electronic wavefunction (with amplitudes $C_1, C_2$):
	\begin{equation}\label{ODE}
	\frac{d}{dt}
	\begin{pmatrix}
	C_1 \\ C_2 
	\end{pmatrix}
	= -\frac{i}{\hbar} 
	\begin{pmatrix}
	H^{el}_{11} & H^{el}_{12} \\ H^{el}_{21} & H^{el}_{22}
	\end{pmatrix}
	\begin{pmatrix}
	C_1 \\ C_2 
	\end{pmatrix}
	\end{equation}
	Here $H^{el}_{ij}$ is a matrix element of the operator $\hH^{el} = \hH_s - \int d\vecr \vE(\vecr)\cdot\hP(\vecr)$.
	
	Second, under the dynamics governed by Eqns. (\ref{density_matrix_evolution}) and (\ref{Maxwell}), the total energy of the system $U_{tot}$ is conserved, where 
	\begin{equation}
	U_{tot} = \frac{1}{2} \int d\vecr \left ( \epsilon_0 |\vE(\vecr)|^2 + \frac{1}{\mu_0} |\vB(\vecr)|^2 \right ) + \text{Tr}\left( \rho\hH_s \right)
	\end{equation}
	Altogether, Eqns. (\ref{density_matrix_evolution}), (\ref{Maxwell}), and (\ref{current_J}) capture the correct physics such that, when an electron decays from the excited state $\ket{e}$ to the ground state $\ket{g}$, an EM field is generated while the total energy is conserved.

	\subsubsection{Advantages and disadvantages of Ehrenfest dynamics} \label{sec:draw_back_of_ehrenfest}
The main advantage for Ehrenfest dynamics is a consistent, simple approach for simulating electronic and EM dynamics concurrently.
	
	Several drawbacks, however, are also apparent for Ehrenfest dynamics. First, consider Eqn. (\ref{eq:J_simplified}).  Certainly, if the initial electronic state is an eigenstate of $H_s$, i.e. $(C_1, C_2) = (0, 1)$, then $\rho_{12}(t=0) = C_1C_2^{\ast}= 0$ and there will be no current density $\vJ(\vecr)$ if there is no EM field initially in space. Thus, in disagreement with the exact quantum result, there is no spontaneous emission: the initial state $(0, 1)$ will never decay.  According to Ehrenfest dynamics, spontaneous emission can be observed only if  $C_1\neq 0$ and $C_2\neq 0$, i.e., if the initial state is a  linear combination of the ground and excited states.

	Second, it is well known that, for finite temperature, Ehrenfest dynamics predicts incorrect electronic populations at long time: the electronic populations will not satisfy detailed balance\cite{Parandekar2006}. Here, finite temperature would correspond to a thermal distribution of photon modes at time $t=0$, representing the black-body radiation. However, for the purposes of fast absorption and/or scattering experiments, where there is no equilibration, this failure may not be fatal.

	\subsection{The Classical Path Approximation (CPA)}
	If Ehrenfest dynamics provides enough accuracy for a given simulation, the relevant dynamics can actually be further simplified and reduced to the standard ``classical path approximation (CPA)''\cite{Smith1969}.  To make this reduction, note that the EM field can be considered the sum of 2 parts: $(i)$ the external EM field $ \vE_{\text{ext}}(\vecr)$ that represents a pulse of light approaching the electronic system and $(ii)$ the 
	scattered EM field $\vE_{\text{scatt}}(\vecr)$ generated from spontaneous or stimulated emission from the molecule itself. Thus, at any time, $\vE(\vecr) = \vE_{\text{ext}}(\vecr) + \vE_{\text{scatt}}(\vecr)$, where we impose
	free propagation for the external EM field, i.e., $\vE_{\text{ext}}(\vecr,t) = \vE_{\text{ext}}(\vecr-ct\hat{r}_{\text{ext}},0)$. Here $\hat{r}_{\text{ext}}$ represents the unit vector in the propagation direction of the external EM field.

	According to the CPA, we ignore any feedback from electronic evolution upon the EM field, i.e., we neglect the $\int d\vecr \ {\vE_{\text{scatt}}(\vecr)} \cdot \hP(\vecr)$ term of Eqn. (\ref{density_matrix_evolution}). Thus, the electronic dynamics now obey
	\begin{equation}\label{density_matrix_reduced}
	\frac{d}{dt}\hrho(t) = -\frac{i}{\hbar}[\hH_s - \int d\vecr \ \vE_{\text{ext}}(\vecr-ct\hat{r}_{\text{ext}})\cdot \hP(\vecr),\  \hrho(t) ]
	\end{equation}
	while photon dynamics still obeys Eqn. (\ref{Maxwell}). This so called classical path approximation underlines all usual descriptions of linear spectroscopy, and should be valid when $|\vE_{\text{scatt}}| \ll |\vE_{\text{ext}}|$. 
	In such a case, the coherence $\rho_{12}$ and current density $\vJ$ are almost unchanged if we neglect the $\int d\vecr \ \vE_{\text{scatt}}(\vecr)\cdot \hP(\vecr)$ term.

	\subsubsection{Advantages and disadvantages of the CPA}
	Obviously, the advantage of Eqn. (\ref{density_matrix_reduced}) over Eqn. (\ref{density_matrix_evolution}) is that we can write down an analytical form for the light-matter coupling ($\int d\vecr E(\vecr)P(\vecr)$), since $\vE_{\text{ext}}$ propagates freely.

	That being said, the disadvantage of the CPA is that one cannot obtain a consistent description of spontaneous emission for the electronic degrees of freedom, because the total energy is not conserved; see Eqns. (\ref{Maxwell}) and (\ref{density_matrix_reduced}). As such, the classical path approximation would appear reasonably only for studying the electronic dynamics; EM dynamics are reliable only for short times.

	\subsection{Symmetrical Quasi-classical (SQC) Windowing Method}	
	As discussed above, the Ehrenfest approach cannot predict exponential decay (i.e. spontaneous emission) when the initial electronic state is $(0, 1)$.  Now, if we want to model spontaneous emission, the usual approach would be to include the vacuum fluctuations of the electric field, in the spirit of stochastic electrodynamics\cite{de2013quantum}. That being said, however, there are other flavors of mean-field dynamics which can improve upon Ehrenfest dynamics and fix up some failures.\cite{Huo2012, kapral:2008:jcp_pbme} (i.e., the inability to achieve branching, the inability to recover detailed balance, etc.)
	Miller's symmetrical quasi-classical (SQC) windowing\cite{Cotton2013}  is one such approach.

	The basic idea of the SQC method is to propagate Ehrenfest-like trajectories with quantum electrons and classical photons (EM field), assuming two modifications: (a) one converts each electronic state to a harmonic oscillator and includes the zero point energy (ZPE) for each electronic degree of freedom (so that one samples many initial electronic configurations and achieves branching); and (b) one bins the initial and final electronic states symmetrically (so as to achieve detailed balance).  We note that SQC dynamics is based upon the original Meyer-Miller transformation\cite{Meyera1979}, which was formalized by Stock and Thoss\cite{Stock1997}, and that there are quite a few similar algorithms that propagate Ehrenfest dynamics with zero-point electronic energy\cite{kapral:2008:jcp_pbme}.	
	While Cotton and Miller have usually propagated dynamics either in action-angle variables or Cartesian variables,
	for our purposes we will propagate the complex amplitude variable $C_1, C_2$ so as to make easier contact with Ehrenfest dynamics\cite{Bellonzi2016}. Formally, $C_j = (x_j + ip_j)/\sqrt{2}$, where $x_j$ and $p_j$ are the dimensionless position and momentum of the classical oscillator.
	
	For completeness,  we will now briefly review the nuts and bolts of the SQC method for a  two-level system coupled to a bath of bosons. 
	
	\subsubsection{Standard SQC procedure for a two-level system coupled to a EM field}
	
	1. At time $t=0$,  the initial complex amplitudes $C_1(0)$ and $C_2(0)$ are generated by Eqn. (\ref{eq:C_to_n_theta}),
	\begin{equation}\label{eq:C_to_n_theta}
	C_j(0) = \sqrt{n_j + \gamma\cdot \text{RN}} \cdot e^{i\theta_j} \ \ \ \ j=1, 2
	\end{equation}
	Here,  $\text{RN}$ is a random number  distributed uniformly  between $[0, 1]$ and $n_j = 0, 1$ is the action variable for electronic state $j$. $n_j = 0$ implies that state $j$ is unoccupied  while $n_j = 1$ implies state $j$ is occupied. $\theta_j=2\pi \text{RN}$ is the angle variable for electronic state $j$. Note that $|C_1|^2 + |C_2|^2 \neq 1$, but rather, on average $|C_1|^2 + |C_2|^2 = 1+2\gamma$, such that $\gamma$ is a parameter that reflects the amount of  zero point energy (ZPE) included. Originally, $\gamma$ was derived to be $1/2$\cite{Meyera1979}, but Stock \textit{et al.} \cite{Stock1995} and Cotton and Miller\cite{Cotton2013} have found empirically that $0 < \gamma <1/2$  often gives better results. 
	
	2. The amplitudes $(C_1,C_2)$ and the field $E,B$ are propagated simultaneously by integrating Eqns. (\ref{ODE}) and (\ref{Maxwell}). 

	3. For each trajectory, transform the complex amplitudes  to action-angle variables according to Eqn. (\ref{eq:nq_to_C})
	\begin{eqnarray}\label{eq:nq_to_C}
	\begin{split}
	n_j &= |C_j|^2 - \gamma \\
	\theta_j &= \tan^{-1}\left ( \frac{\text{Im}C_j}{\text{Re}C_j} \right ) \text{\ \ \ \ } j= 1,2
	\end{split}
	\end{eqnarray}
	
	4. At each time $t$, one may calculate raw populations (before normalization)  as follows:
	\begin{eqnarray}\label{raw_population}
	\begin{split}
	\tilde{P}_1(t) &= \sum_{l=1}^{N} W_2(\mathbf{n}^{(l)}, \mathbf{q}^{(l)}, t=0)W_1(\mathbf{n}^{(l)}, \mathbf{q}^{(l)}, t) \\
	\tilde{P}_2(t) &= \sum_{l=1}^{N} W_2(\mathbf{n}^{(l)}, \mathbf{q}^{(l)}, t=0)W_2(\mathbf{n}^{(l)}, \mathbf{q}^{(l)}, t)
	\end{split}
	\end{eqnarray}
	Here, $N$ is the number of trajectories and $W_1$ is the window function for the  ground state $\ket{g}$, centered at $(n_1, n_2) = (1, 0)$; $W_2$ is the window function for the excited state $\ket{e}$, centered at $(n_1, n_2)  = (0, 1)$.
	$(l)$ means the $l$th trajectory. 

	5. The true density matrix at time $t$ is calculated by normalizing Eqn. (\ref{raw_population}) in the following manner:
	\begin{subequations}\label{true_population}
	\begin{align}
	P_1(t) &= \frac{	\tilde{P}_1(t)}{\tilde{P}_1(t) + \tilde{P}_2(t)} \\
	P_2(t) &= \frac{	\tilde{P}_2(t)}{\tilde{P}_1(t) + \tilde{P}_2(t)} \label{eq:Pj}
	\end{align}
	\end{subequations}
	Miller and Cotton have also proposed a protocol to calculate coherences and not just populations\cite{Miller2016}, but we have so far been unable to extract meaningful values from this approach. Future work exploring such coherences would be very interesting.

	\subsubsection{Choice of window function and initial distribution}
\begin{figure}
		\includegraphics[width=8.5cm]{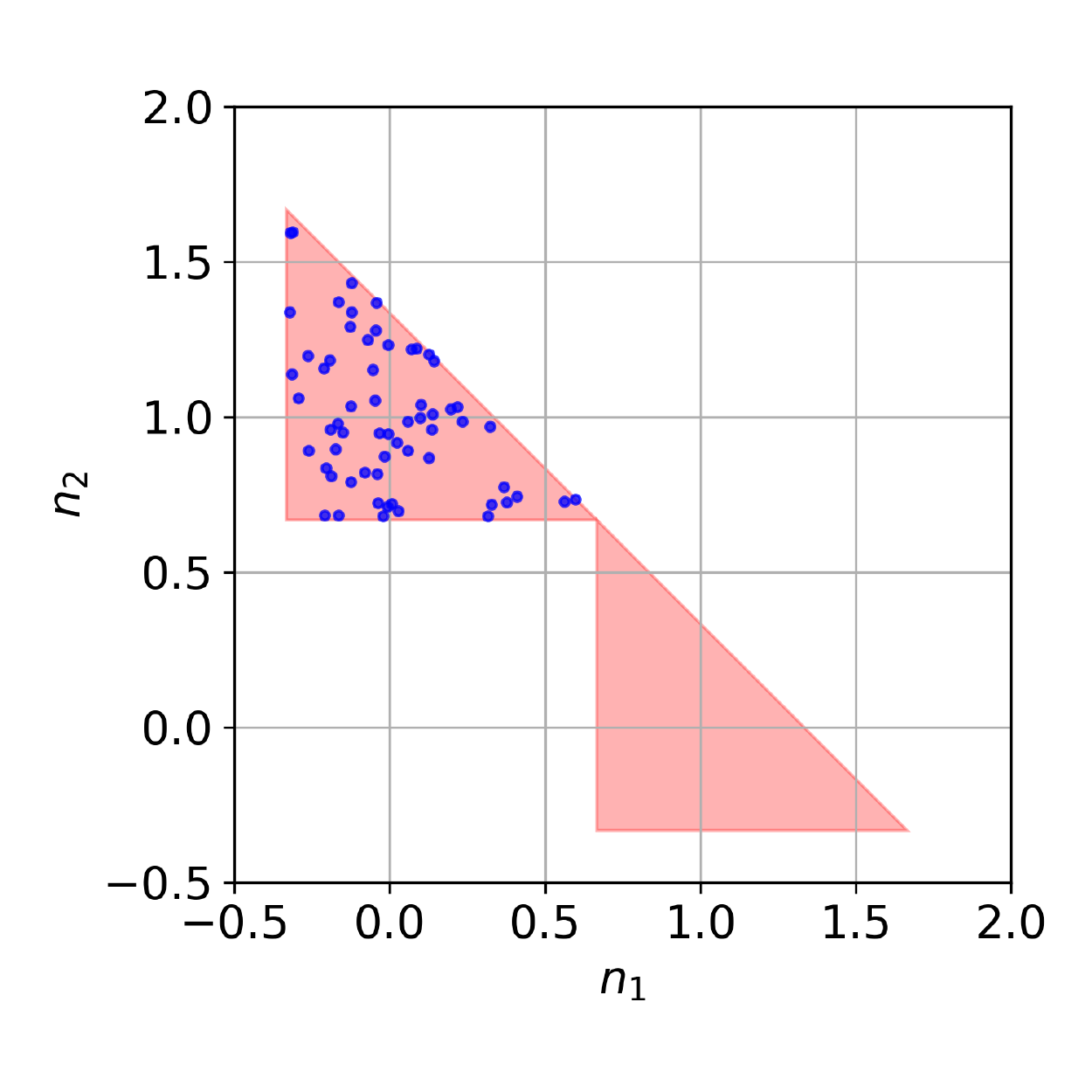}
		\caption{\label{fig:sqc_demo} A plot of the initial $(n_1, n_2)$ distribution as required by the SQC algorithm.  The upper and lower pink triangles represents areas where the triangular window function $W_2\neq 0$ and $W_1\neq 0$,  respectively; see  Eqn. \ref{eq:triangle_bin} . The initial values of $(n_1, n_2)$ (blue dots) are uniformly distributed within the upper triangular area ($W_2\neq 0$).}
	\end{figure}

	Below, we will study a two-level system weakly coupled to the EM field, i.e.  the polarization energy will be several orders less than $\hbar\omega_0$.  For such a case, one must be very careful about binning. Cotton and Miller \cite{Cotton2016} have suggested that triangular window functions with $\gamma=1/3$ perform better than  square window functions in this regime. 
	Therefore, we have invoked the triangular window function in Eqn. (\ref{eq:triangle_bin}) with $\gamma=1/3$ below.
	\begin{eqnarray}\label{eq:triangle_bin}
	\begin{split}
	W_1(n_1, n_2) = & 2\cdot h(n_1 + \gamma - 1)\cdot h(n_2+\gamma) \\ 
	& \times h(2-2\gamma-n_1-n_2)\\
	W_2(n_1, n_2) = & 2\cdot h(n_1 + \gamma)\cdot h(n_2+\gamma-1) \\ 
	& \times h(2-2\gamma-n_1-n_2)
	\end{split}
	\end{eqnarray}
	Here, $h(x)$ is Heaviside function.   Fig. \ref{fig:sqc_demo} gives a visual representation of the triangular window function in Eqn. (\ref{eq:triangle_bin}). The bottom and upper pink triangles represent areas where $W_1\neq 0$ and $W_2\neq 0$  respectively.
	
	To be consistent with the choice of triangular window functions, one must modify the standard  protocol in Eqn. (\ref{eq:C_to_n_theta}). 
	Instead of the standard square protocol, assuming we start in excited state $\ket{e}$, one generates a distribution of initial action variables $(n_1(0), n_2(0))$ within the area where $W_2 \neq 0$ (see Eqn. \ref{eq:triangle_bin}) uniformly. Visually, this initialization implies  a distribution of $(n_1(0), n_2(0))$ inside a triangle centered at $(0, 1)$ in the $(n_1, n_2)$ configuration space, as demonstrated in Fig. \ref{fig:sqc_demo}. The protocol for initializing angle variables is not altered: one sets  $\theta_j = 2\pi \text{RN}, \ j=1,2$.

	\subsubsection{Advantages and disadvantages of SQC dynamics}
	Compared with Ehrenfest dynamics, one obvious advantage of SQC dynamics is that the latter can model spontaneous emission when the initial electronic state is $(0, 1)$. Moreover, the SQC approach must recover detailed balance in the presence of a photonic bath at a given temperature\cite{miller2015communication} --- provided that the parameter $\gamma$ is chosen to be small enough for the binning\cite{Bellonzi2016}.
	
	At the same time, the disadvantage of the SQC method is that all results are sensitive to the binning width $\gamma$. $\gamma$ should be big enough to give enough branching, but also should be small enough to enforce detailed balance\cite{Bellonzi2016}. As a result, one must be careful when choosing $\gamma$.  Although not relevant here, it is also true that SQC can be unstable for anharmonic potentials.\cite{Bellonzi2016} Lastly, as a practical matter, we have found SQC requires about $1000$ times more trajectories than Ehrenfest dynamics.
	
	\subsection{Classical Dynamics with Abraham-Lorentz Forces}
	Although (as shown above) classical electrodynamics with Abraham-Lorentz forces can be useful to model self-interaction, we will not analyze Abraham-Lorentz dynamics further in this paper. Because the correspondence between Ehrenfest dynamics and Abraham-Lorentz dynamics is not unique or generalizable, we feel any further explanation of Abraham-Lorentz equation would be premature. While a Meyer-Miller transformation\cite{Meyera1979} can reduce a quantum mechanical Hamiltonian into a classical Hamiltonian, the inverse is not possible. Thus, it is not clear how to run classical dynamics with Abraham-Lorentz forces starting from an arbitrary initial superposition state $(C_1, C_2)$ in the $\{ \ket{g}, \ket{e}\}$ basis. 
	For instance, following the approach above in Section \ref{sec:AL}, we might set $m\omega_0^2 \left \langle x^2\right \rangle = |C_2(0)|^2 \hbar \omega_0/2$.  However, doing so leads to a rate of decay equal to $k_{\text{FGR}}/|C_2(0)|^2$.  This result goes to infinity in the limit  $C_2 \rightarrow 0$; see Fig. \ref{fig:pulse_1d_k_p2_relation}.  Future work  may succeed at finding the best correspondence between semiclassical dynamics and the Abraham-Lorentz framework, but such questions will not be the focus of the present paper.
	
	\section{Simulation Details}\label{sec:simulation_details}
	\subsection{Parameter Regimes}
	We focus below on Hamiltonians with electronic dipole moment $\mu_{12}$  in the range of $2000 \sim 50000$ C$\cdot$nm/mol ($1 \sim 25$ in Debye) and electronic energy gaps $\hbar \omega_0$  in the range of $3 \sim 25 $ eV. Other practical parameters are chosen as in Table \ref{tab:table1}.  Two different sets of simulations are run: $(i)$ simulations to capture spontaneous emission (with zero EM field initially) and $(ii)$ simulations to capture stimulated emission (with an incoming external finite EM pulse located far away at time zero).
	
	\begin{table}
		\caption{\label{tab:table1} Default Numerical Parameters. $N_{\text{grids}}$ is the number of grid points in each dimension for the EM field. $X_{\text{max}}$ and $X_{\text{min}}$ are the boundary points in each dimension. $dt$ and $t_{\text{max}}$ are the time step and maximum time of simulation respectively. ABC denotes ``Absorbing Boundary Conditions''.}
		\begin{ruledtabular}
			\begin{tabular}{lccr}
				Quantity&1D no ABC &1D with ABC& 3D with ABC\\
				\hline
				$\hbar \omega_0$\footnote{Eqn. (\ref{eq:Hs})} (eV)& 16.46 & 16.46 & 16.46\\
				$\mu_{12}$\footnote{Eqns. (\ref{eq:3dxi}, \ref{eq:1dxi})} (C$\cdot$nm/mol)\footnote{As mentioned before, $\mu_{12}$ has dimension of C/mol in 1D and C$\cdot$nm/mol in 3D}& 11282& 11282 & 23917\\
				$a$\footnote{Eqns. (\ref{eq:3dxi}, \ref{eq:1dxi})} ($\text{nm}^{-2}$)& 0.0556& 0.0556 & 0.0556\\
				$N_{\text{grids}}$& 40000& 200 & 60\\
				$X_{\text{max}}$ (nm) & 2998 & 89.94& 89.94\\
				$X_{\text{min}}$(nm) & -2998 & -89.94 & -89.94\\
				$dt$ (fs) &$2\times10^{-4}$ & $2\times10^{-4}$ & $5\times10^{-4}$\\
				$t_{\text{max}}$ (fs) & 99& 99 & 500\\
				$R_0$\footnote{Eqns. (\ref{eq:smooth_function_1d}-\ref{eq:smooth_function_3d})} (nm)& -&50& 50\\
				$R_1$\footnote{Eqns. (\ref{eq:smooth_function_1d}-\ref{eq:smooth_function_3d})} (nm)& -& 84&84\\
			\end{tabular}
		\end{ruledtabular}
	\end{table}
	
	\subsection{Propagation procedure}
	
	Equations of motion (Eqns. (\ref{density_matrix_evolution}), (\ref{Maxwell})) are propagated with a Runge-Kutta 4th order  solver, and all spatial gradients are evaluated on a real space grid with a two-stencil in 1D and a six-stencil in 3D. Thus, for example, if we consider Eqn. (\ref{Maxwell}) in 1D, in practice we approximate:
	\begin{eqnarray}\label{eq:1d_em_numerical}
	\begin{split}
	\frac{d B_y^{(i)}}{dt} &=  \frac{E_z^{(i+1) }- E_z^{(i-1)}}{2\Delta r} \\
	\frac{d E_z^{(i)}}{dt} &=  c^2\frac{B_y^{(i+1)} - B_y^{(i-1)}}{2\Delta r}  - \frac{J_z^{(i)}}{\epsilon_0},
	\end{split}
	\end{eqnarray}
	etc. Here $(i)$ is a grid index. This numerical method to propagate the EM field (Eqn. (\ref{Maxwell})) is effectively a  \textit{finite-difference time-domain }(FDTD) method\cite{taflove1998advances, harris2011computational}. 
	
	\subsection{Absorbing boundary condition (ABC)}  
	To run calculations in 3D, absorbing boundary condition (ABC) are required to alleviate the large computational cost. For such a purpose, we invoke a standard, one-dimensional smoothing function\cite{Subotnik2005, Subotnik2006} $S(x)$:
	\begin{equation}\label{eq:smooth_function_1d}
	S(x)=
	\begin{cases}
	\hphantom{-} 1 &\text{ $|x| < R_0$}, \\[2ex]
	\hphantom{-} \left[ 1 + e^{-\left ( \frac{R_0-R_1}{R_0 -|x|} +  \frac{R_1-R_0}{|x| -R_1} \right ) } \right ]^{-1} &\text{$R_0 \leq |x| \leq R_1$}, \\[2ex]
	\hphantom{-} 0 &\text{$|x| > R_1$}
	\end{cases}
	\end{equation}
	In 1D, by multiplying the E and B field with $S(x)$ after each time step, we force the E and B fields to vanish for $|x| > R_1$.

	In 3D, we choose the corresponding smoothing function to be of the form of Eqn. (\ref{eq:smooth_function_3d}),
	\begin{equation}\label{eq:smooth_function_3d}
	S(\vecr) = S(x)S(y)S(z)
	\end{equation}
	where $S(x)$, $S(y)$ or $S(z)$ is exactly the same as Eqn. (\ref{eq:smooth_function_1d}). Note that this 
	smoothing function has cubic (rather than spherical)  symmetry. 
	
	For the simulations reported below, applying ABC's allows us to keep only $\sim 1\%$  of the grid points in each dimension, so that the computational time is reduced by a factor of $10^2$ in 1D and by a factor of $10^6$ in 3D.
	Our use of ABC's is benchmarked in Figs. \ref{fig:ehrenfest_1d_1}-\ref{fig:ehrenfest_1d_2}, and ABC's are used implicitly for SQC dynamics in Figs. \ref{fig:sqc_1d_2},  \ref{fig:pulse_1d_U_4000}, \ref{fig:pulse_1d_k_p2_relation} and \ref{fig:dump_1d}. ABC's are also used for the 3D dynamics in Fig. \ref{fig:ehrenfest_3d_2}.
	
	\subsection{Extracting Rates}\label{sec:default_parameters} 
	
	Our focus below will be on calculating rates of emission; these rates will be subsequently compared with FGR rates. To extract a numerical rate $(k)$  from Ehrenfest or SQC dynamics, we  simply calculate the probability to be on the excited state as a function of time  ($P_2(t)$) and fit that probability to an exponential decay: $P_2(t) \equiv P_2(0)e^{-kt}$. For Ehrenfest dynamics, all results are converged using the default parameters in  Table \ref{tab:table1}. 
	For SQC dynamics,  longer simulation times are needed (to ensure $P_2(t_{\text{end}})<0.02$);
	in practice, we set $t_{\text{end}}$ = 150 fs. Note that, for SQC dynamics, $P_2(t)$ in SQC is calculated by Eqn. (\ref{eq:Pj}) and we sample 2000 trajectories. 

	\section{Results}\label{sec:result}
	We now present the results of our simulations and analyze how Ehrenfest and SQC dynamics treat
	spontaneous emission. The initial state is chosen to be $(C_1, C_2)$ = $(\sqrt{1/2}, \sqrt{1/2})$ for Ehrenfest dynamics. We begin in one-dimension.

	\subsection{Ehrenfest Dynamics:  1D }
\begin{figure}
		\includegraphics[width=9cm]{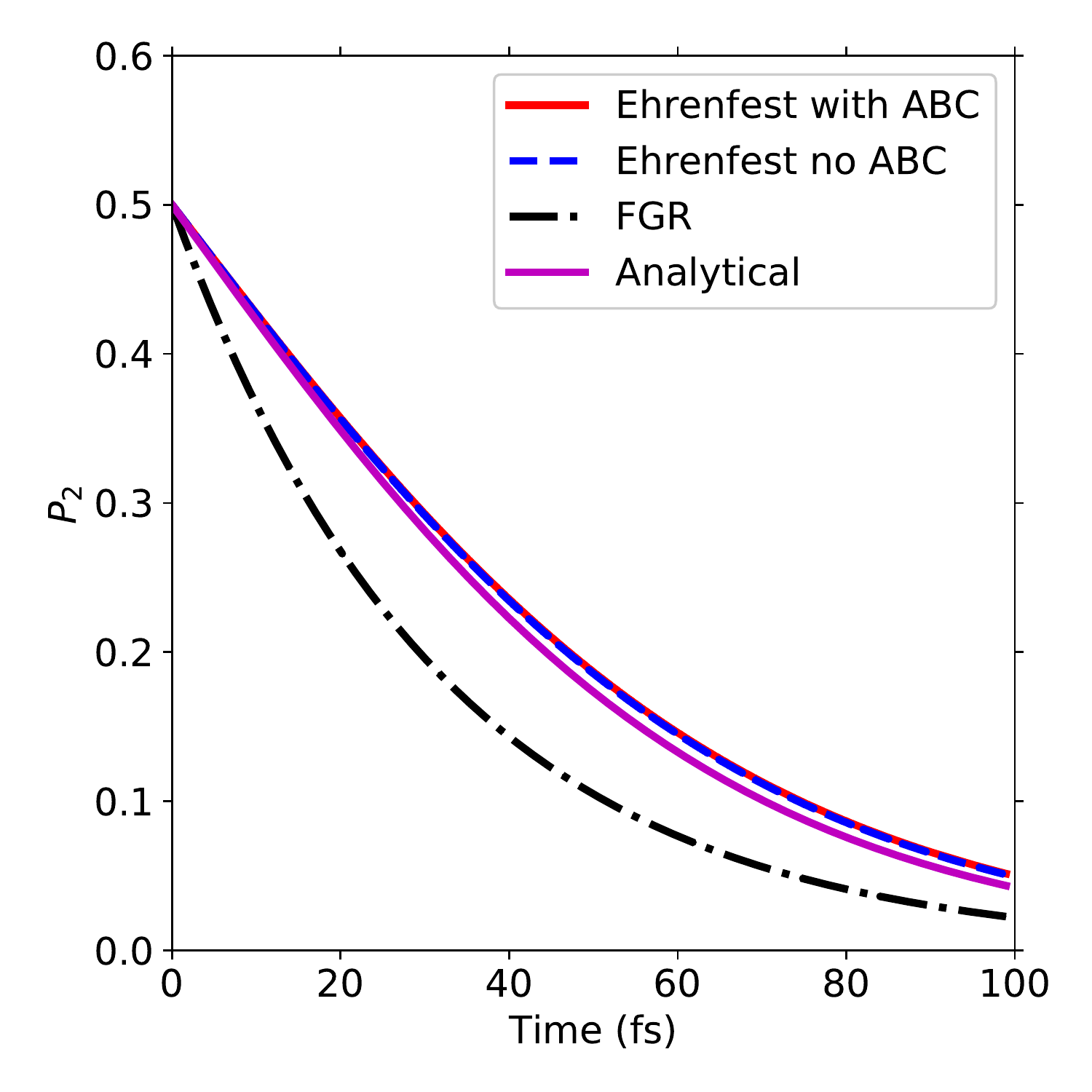}
		\caption{\label{fig:ehrenfest_1d_1} Spontaneous decay rate according to Ehrenfest dynamics in 1D.  Here, we plot the electronic population in the excited state $\ket{e}$, $P_2$, as a function of time $t$ using the default parameters in Table \ref{tab:table1}. The initial electronic state  is $(|C_1|, |C_2|)$ = $(\sqrt{1/2}, \sqrt{1/2})$. The results do not depend on the initial phases of $C_1$ and $C_2$. The analytical Ehrenfest result (magenta line) is plotted according to Eqn. (\ref{eq:ehrenfest_analytical}) in Appendix A.} 
	\end{figure}

	In Fig. \ref{fig:ehrenfest_1d_1}, we plot $P_2(t)$ for the  default parameters in Table \ref{tab:table1}. Clearly, including ABC's has no effect on our results. For this set of parameters,  Ehrenfest dynamics predicts a decay rate that is  $\sim1/3$ slower than Fermi's Golden Rule (FGR) in Eqn. (\ref{eq:FGR_1d}). 

	In Fig. \ref{fig:ehrenfest_1d_2}, we now examine the behavior of Ehrenfest dynamics across a broader parameter regime. In Fig. \ref{fig:ehrenfest_1d_2}\textit{a} and \ref{fig:ehrenfest_1d_2}\textit{b}, we plot the dependence of the decay rate on the energy difference of electronic states, $\hbar\omega_0$, and the dipole moment, $\mu_{12}$.   Ehrenfest dynamics correctly predicts linear and quadratic dependence, respectively, in agreement with FGR in 1D (see Eqn. (\ref{eq:FGR_1d})).   Generally, the fitted decay rate from Ehrenfest dynamics is $\sim 1/3$ slower than FGR.  As far as the size of the molecule is concerned,  in Fig. \ref{fig:ehrenfest_1d_2}\textit{c}, we plot the decay rate $k$ as a function of the parameter $a$ (in Eqn. \ref{eq:1dxi}).  Note that our results are independent of molecular size when $a > 0.05$ nm$^{-2}$.  This independence underlies the dipole approximation: when the width of the molecule is much smaller than  wavelength of light, $ \sqrt{1/a} \ll c/\omega_0 $, the decay rate should not be dependent on the width of molecule.  Note that  $\hbar \omega_0=16.46$ eV for these simulations, which dictates that results will be dependent on $a$ for $a < 0.05$ nm$^{-2}$. Finally,  Fig. \ref{fig:ehrenfest_1d_2}\textit{d} should convince the reader that our decay rates are converged with the density of grid points.

	\begin{figure}
		\includegraphics[width=9cm]{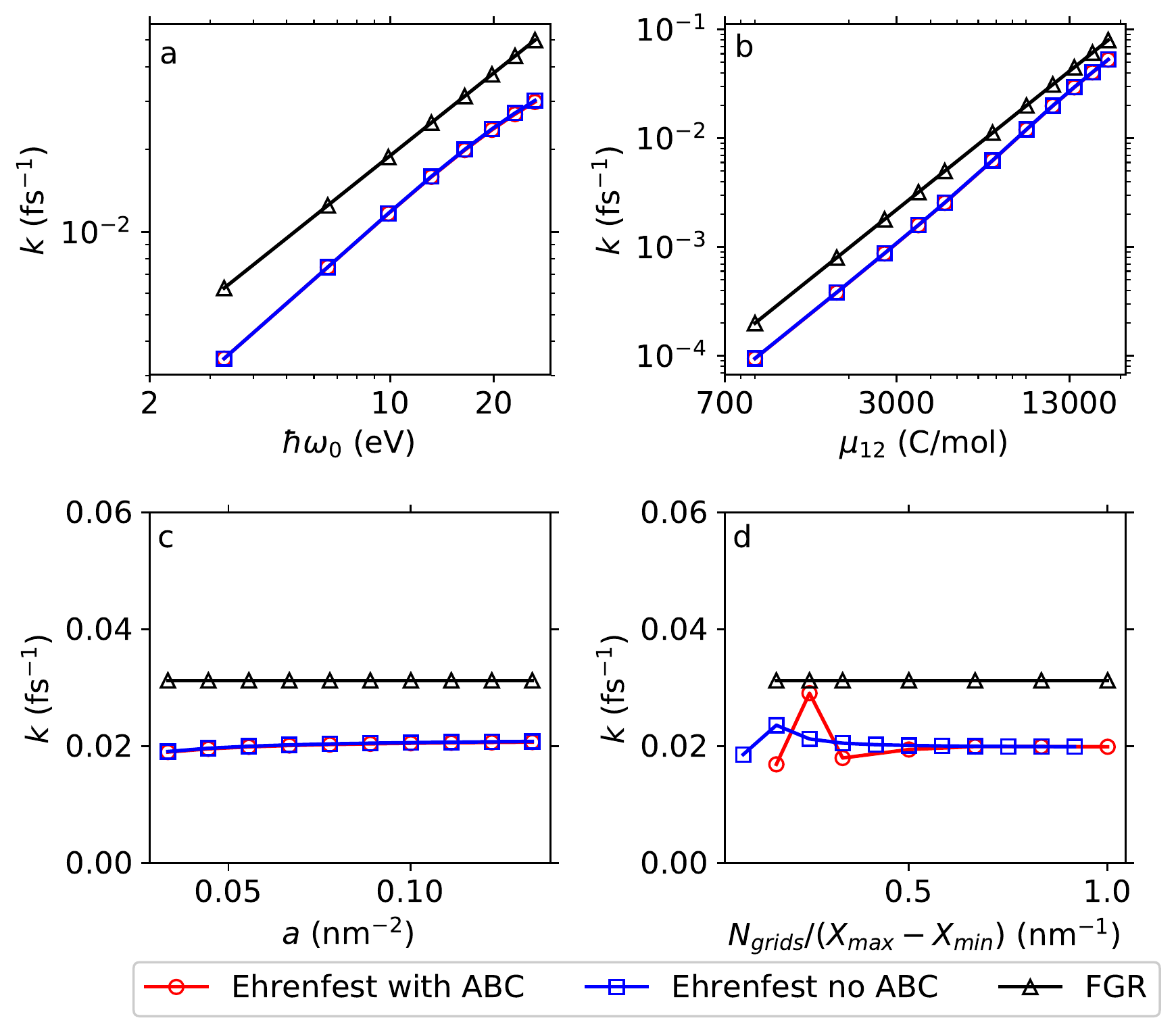}
		\caption{\label{fig:ehrenfest_1d_2} Analyzing the dependence of Ehrenfest spontaneous decay on the system variables in 1D. Here we plot the fitted decay rate $k$ versus (\textit{a}) the  energy difference between electronic states, $\hbar\omega_0$; (\textit{b}) the electronic transition dipole moment $\mu_{12}$; (\textit{c}) the Gaussian width parameter $a$; (\textit{d}) the density of  $N_{\text{grids}}$. Three approaches are compared: Ehrenfest dynamics with ABC (red $\circ$), Ehrenfest dynamics without ABC (blue $\Box$) and Fermi's Golden Rule (black $\bigtriangleup$). Extraneous parameters are always set to their default values in Table \ref{tab:table1}. The initial electronic state is $(C_1, C_2)$ = $(\sqrt{1/2}, \sqrt{1/2})$.  Note that Ehrenfest dynamics captures  most of the correct FGR physics.
		}
	\end{figure}

	\subsubsection{Initial Conditions}
\begin{figure}
		\includegraphics[width=8cm]{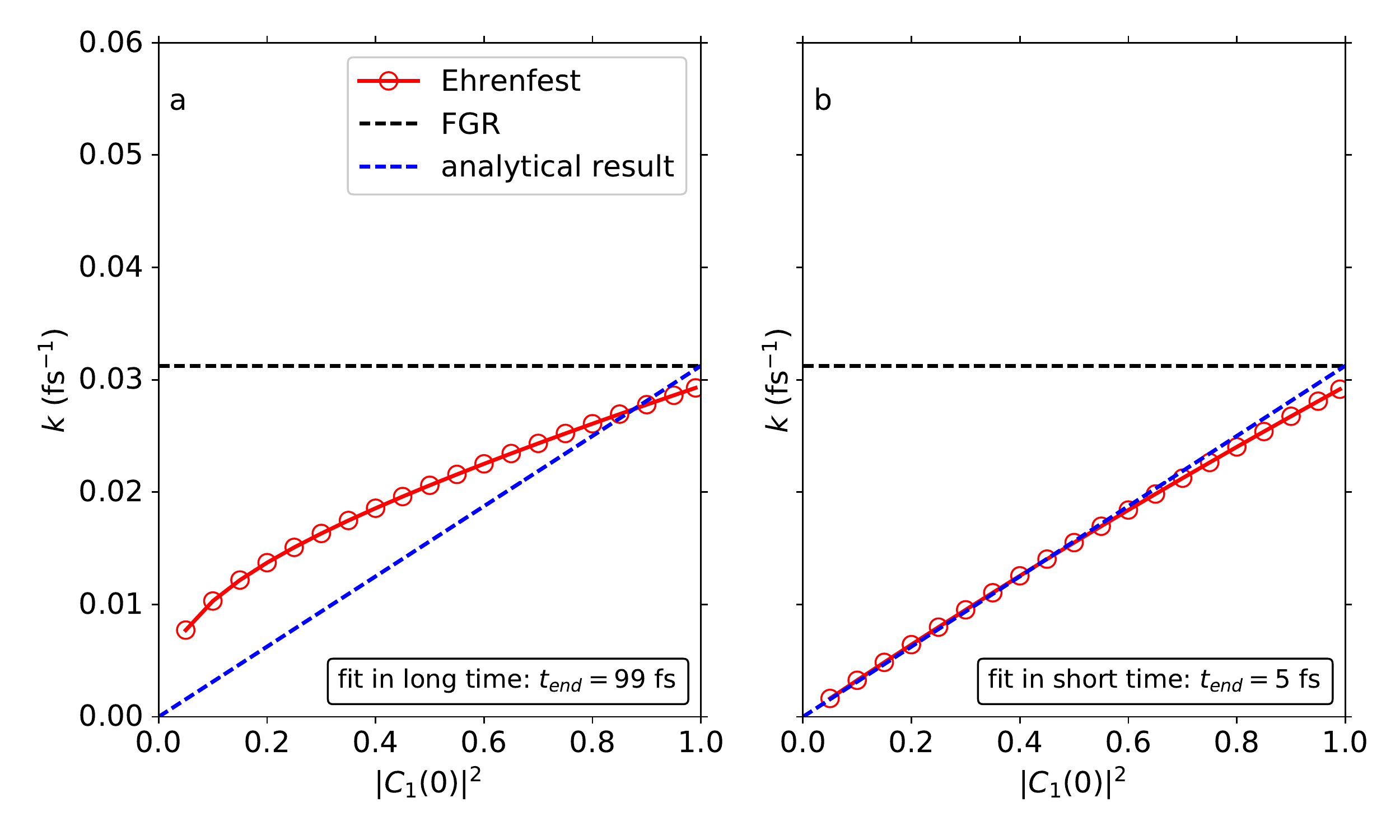}
		\caption{\label{fig:ehrenfest_1d_c1} The dependence of the 1D Ehrenfest spontaneous decay rate  ($k$) as a function of  the initial population on the ground state $|C_1(0)|^2$.  Note that the decay is not purely exponential and depends on whether we invoke  (\textit{a}) a long time fit ($t_{\text{end}} = 99$ fs) or (\textit{b}) a short time fit ($t_{\text{end}} = 5$ fs). Other parameters are set to their default values in Table \ref{tab:table1}. Three approaches are compared: FGR (dashed black), Ehrenfest (red $\circ$) and the analytical, short time result obtained in Appendix, i.e. $k = k_{\text{FGR}}|C_1|^2$ (dashed blue). Note that the analytical result matches up well with the extracted fit in $(b)$. 
		}
	\end{figure}

	The results above were gathered by setting $C_1 =\sqrt{1/2}.$ Let us now address how the initial conditions affect the Ehrenfest rate of spontaneous decay.  In Fig. \ref{fig:ehrenfest_1d_c1} we plot $k$ vs. $|C_1(0)|^2$. Here, we differentiate how $k$ is extracted, either from a $(a)$ a fit of the long time decay  ($t_{\text{end}} = 99$ fs) or $(b)$ a fit of the short time decay  ($t_{\text{end}} = 5$ fs). 
Clearly, the decay rates in Fig. \ref{fig:ehrenfest_1d_c1}\textit{a} and \ref{fig:ehrenfest_1d_c1}\textit{b} are different, suggesting that the decay of $P_2$ is 
not purely exponential (see detailed discussion in Appendix); the decay constant is itself a function of time.  Moreover, according to Fig.
\ref{fig:ehrenfest_1d_c1}$b$, the short time decay rate appears to be linearly dependent  on
$|C_1(0)|^2$ and,  in the limit that $|C_1(0)|^2\rightarrow 1$, both fitted decay rates $k$ approach the FGR result. These results suggest that the fitted decay rate $k$ satisfies
	\begin{equation}\label{eq:c1}
	k = k_{\text{FGR}}|C_1(0)|^2
	\end{equation}
 where $k_{\text{FGR}}$ is the FGR decay rate. In fact, in the Appendix, we will show that Eqn. (\ref{eq:c1}) can be derived for early time scales ($2\pi/\omega_0 \ll  t \ll 1/k_{\text{FGR}}$ ) under certain approximations.  
We also mention that the same failure was observed previously by Tully when investigating the erroneous long time populations predicted by Ehrenfest dynamics.\cite{parandekar2005mixed, parandekar2006detailed, miller2015communication}

	\subsubsection{Distribution of EM field}
			\begin{figure*}
		\includegraphics[width=18cm]{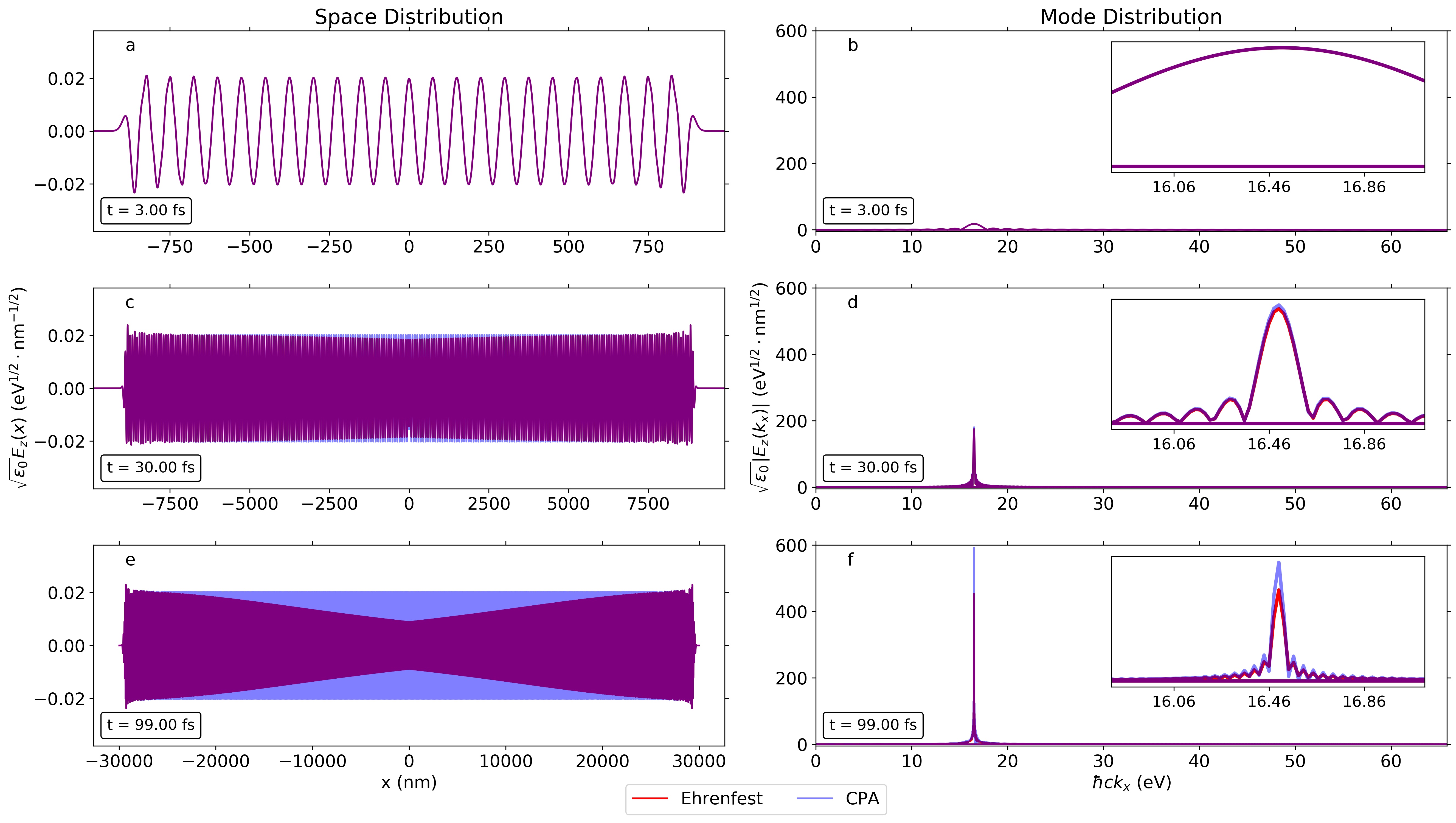}
		\caption{\label{fig:ehrenfest_final_em} An analysis of the EM field produced by spontaneous  emission in 1D.  We plot  (\textit{left}) the distribution of $E_z(x)$ along $x$-axis at times (\textit{a}) 3.00 fs, (\textit{c}) 30.00 fs, (\textit{e}) 99.00 fs and (\textit{right}) the Fourier transform of $E_z(x)$ at the same times. $x$-axis :  the energy of photon modes $\hbar ck_x$; $y$-axis : $\sqrt{\epsilon_0}E_z(k_x)$. The inset figures on the right zoom in on the spectral peaks in the neighborhood of $\hbar\omega_0$ (16.46 eV here). Two Methods are compared: Ehrenfest dynamics (red lines) and the CPA (light blue lines). The default parameters in Table \ref{tab:table1} have been used here. Note that Ehrenfest dynamics and the CPA agree for short times but 
			only Ehrenfest dynamics predicts a decrease in the EM field for larger times, which is a requirement of  energy conservation.}
	\end{figure*}

    Beyond the electronic subsystem, Ehrenfest dynamics allows us to follow the behavior of the EM field directly.
	In Fig. \ref{fig:ehrenfest_final_em}, we plot the  distribution of the EM field at times 3.00 fs (\textit{a-b}) , 30.00 fs (\textit{c-d}), and 99.00 fs (\textit{e-f})  with two methods: Ehrenfest (red lines) and the  CPA (light blue lines). On the left hand side, we plot the electric field in real space ($E_z(x)$); on the right hand side, we plot the EM field in Fourier space ($E_z(k_x)$). Here, the Fourier transform is performed over the region $x>0$, which  corresponds to light traveling exclusively to the right.  In the insets on the right, we zoom in  on the spectra in a small neighborhood of $\hbar\omega_0$ (here, 16.46 eV).

	From Fig. \ref{fig:ehrenfest_final_em}, we find that Ehrenfest dynamics and the CPA agree for short times.
	However, for larger times, only Ehrenfest dynamics predicts a decrease in the EM field (corresponding to the spontaneous decay of the signal).  This decrease is guaranteed by Ehrenfest dynamics because this method conserves energy.   By contrast, because it ignores feedback and violates energy conservation, the CPA does not predict a decrease in the emitted EM field as a function of time (or any spontaneous decay).  Thus, overall, as shown in Fig. \ref{fig:ehrenfest_final_em}\textit{f}, the long time EM signal will be a Lorentzian according to Ehrenfest dynamics or a delta-function according to the CPA.  These conclusions are unchanged for all values of the initial $|C_1(0)|^2$.

	\subsection{SQC: 1D}

\begin{figure}
		\includegraphics[width=9cm]{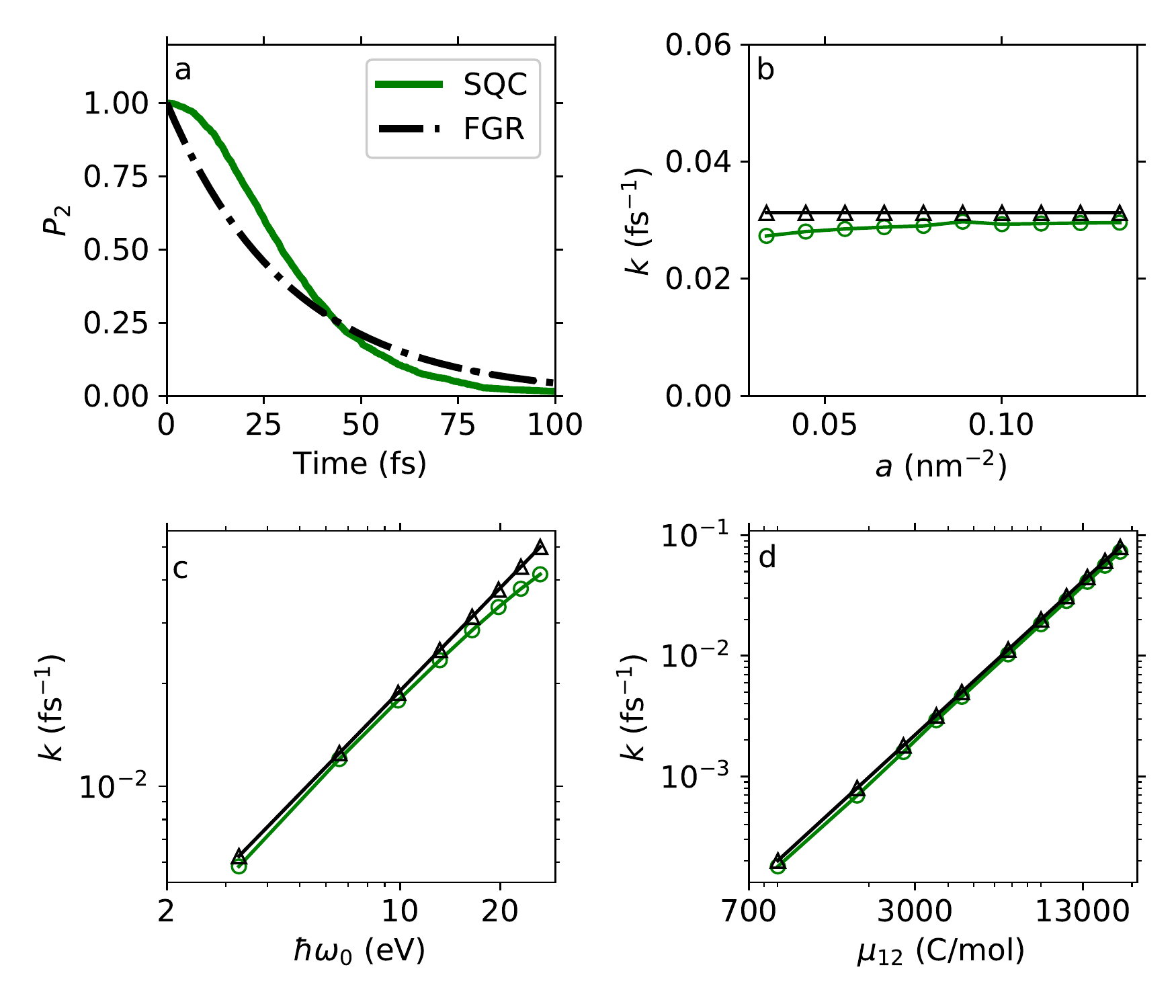}
		\caption{\label{fig:sqc_1d_2} Analysis of SQC spontaneous emission rates in 1D.  In (\textit{a}), we plot the electronic population of the excited state $P_2$ versus time $t$. For the remaining subfigures,  we plot how the  fitted decay rate $k$ depends on (\textit{b}) the Gaussian width parameter $a$, (\textit{c}) the energy difference between the two electronic states $\hbar\omega_0$ and (\textit{d}) the electric transition dipole moment $\mu_{12}$. Two results are compared: SQC dynamics with ABC (Green $\circ$) and Fermi's Golden Rule (black $\bigtriangleup$).  All unreported parameters are set to their default values in Table \ref{tab:table1}. The initial electronic state is $(C_1, C_2) = (0, 1)$. Note that the SQC decay rates are very close to the FGR  rates (less than 10 \% difference), whereas Ehrenfest dynamics completely fail and predicts $k = 0$ for this case (when $C_2 = 1$ initially). For these simulations, we apply ABC's.	
	}

	\end{figure}

	The simulations above have been repeated with SQC dynamics. In Fig. \ref{fig:sqc_1d_2}\textit{a}, we plot $P_2(t)$  for a single trajectory that begins on the excited state  ($C_2 = 1$) for the default parameters (see Table \ref{tab:table1}). The remaining three sub-figures in Fig. \ref{fig:sqc_1d_2} demonstrate the dependence of the fitted decay rate $k$ on (\textit{b}) the molecular width parameter $a$, (\textit{c}) the electronic excited state energy $\hbar\omega_0$ and (\textit{d}) the electronic dipole moment $\mu_{12}$.  Generally, SQC depends on  $a$, $\omega_0$ and $\mu_{12}$ as in a manner similar to Ehrenfest dynamics. However, for the initial condition $C_2 = 1$, the overall SQC decay rate $k$ is almost the same as FGR (less than 10 \% difference), whereas Ehrenfest dynamics completely fails and predicts $k=0$.
	\footnote{For these simulations, we do not consider SQC dynamics as a function of $C_1$ (as in Fig. \ref{fig:ehrenfest_1d_c1}).  In practice, for such simulations, we would need to initialize in one representation and measure in another representation, and thus far,  we have been unable to recover stable data using the techniques in Ref. \onlinecite{Cotton2013}. We believe this failure is likely caused by our own limited experience with SQC.}

	\subsection{Ehrenfest Dynamics: 3D}
		
	\begin{figure}[h]
		\includegraphics[width=9cm]{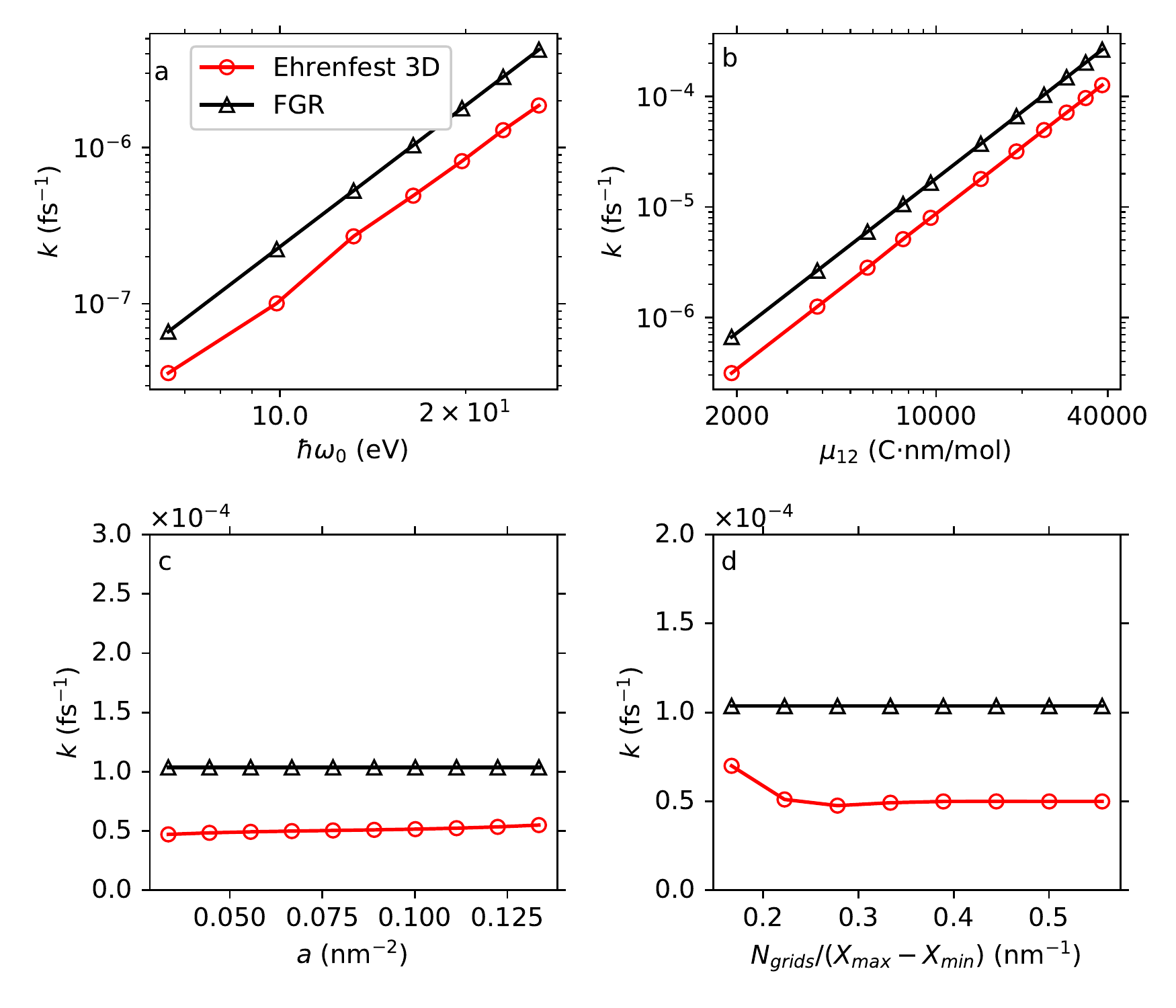}
		\caption{\label{fig:ehrenfest_3d_2} The fitted decay rate $k$ (as predicted by Ehrenfest dynamics in 3D) versus (\textit{a}) the energy difference between electronic states $\hbar \omega_0$; (\textit{b}) the electronic transition dipole moment $\mu_{12}$; and (\textit{c}) the Gaussian width parameter $a$;  and (\textit{d}) the density of grid points $N_{\text{grids}}$ in each dimension. Two results are compared: Ehrenfest dynamics with ABC (red $\circ$) and Fermi's Golden Rule (black $\bigtriangleup$). All unreported parameters are set to their default  as in Table \ref{tab:table1}. The initial electronic state  is $(C_1, C_2)$ = $(\sqrt{1/2}, \sqrt{1/2})$.  The Ehrenfest decay rates in 3D depend correctly only $a$, $\omega_0$ and $\mu_{12}$ and match FGR. For these simulations, we apply ABC's.	}
	\end{figure}
	
	\begin{figure}[h]
	\includegraphics[width=9cm]{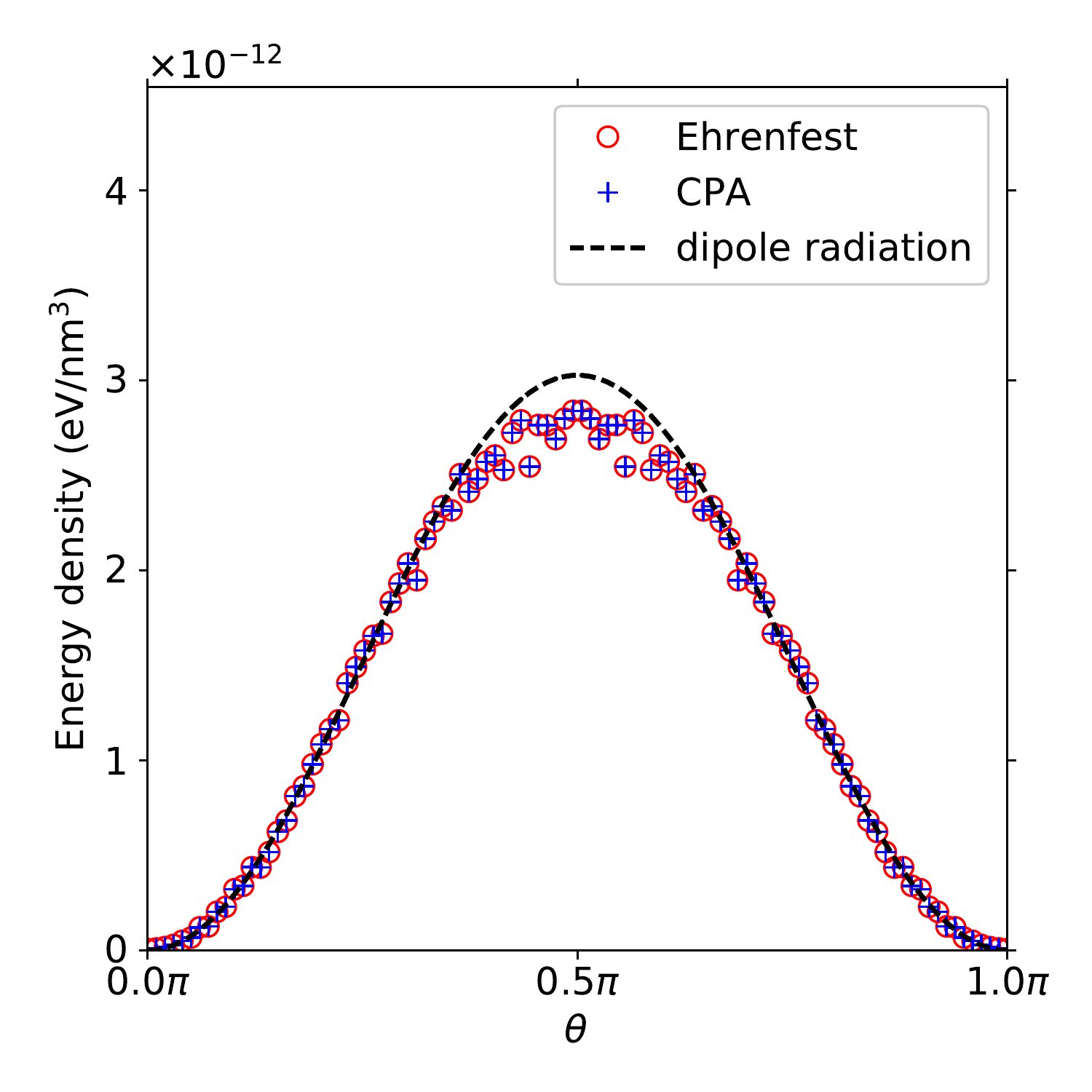}
	\caption{\label{fig:ehrenfest_3d_em}  The energy density of the spontaneous EM field
(as predicted by Ehrenfest dynamics in 3D) versus polar angle $\theta$ when $t=1.00$ fs. Here, all data has  been averaged over a  sphere with $r=294$ nm. The simulation parameters are  $N_{\text{grids}} = 210$, $X_{\text{max}} = 315$ nm and  $X_{\text{min}} = -315$ nm for each dimension.
Unreported parameters are as in Table \ref{tab:table1}. ABCs are not applied here.  The initial electronic state is  $(C_1, C_2)$ = $(\sqrt{1/2}, \sqrt{1/2})$. Note the strong and perhaps surprising agreement between Ehrenfest/CPA  dynamics  and the classical dipole radiation; this agreement depends on the choice of initial electronic states, as is proven in the Appendix.
	}
	\end{figure}
	\begin{figure}[h]
	\includegraphics[width=9cm]{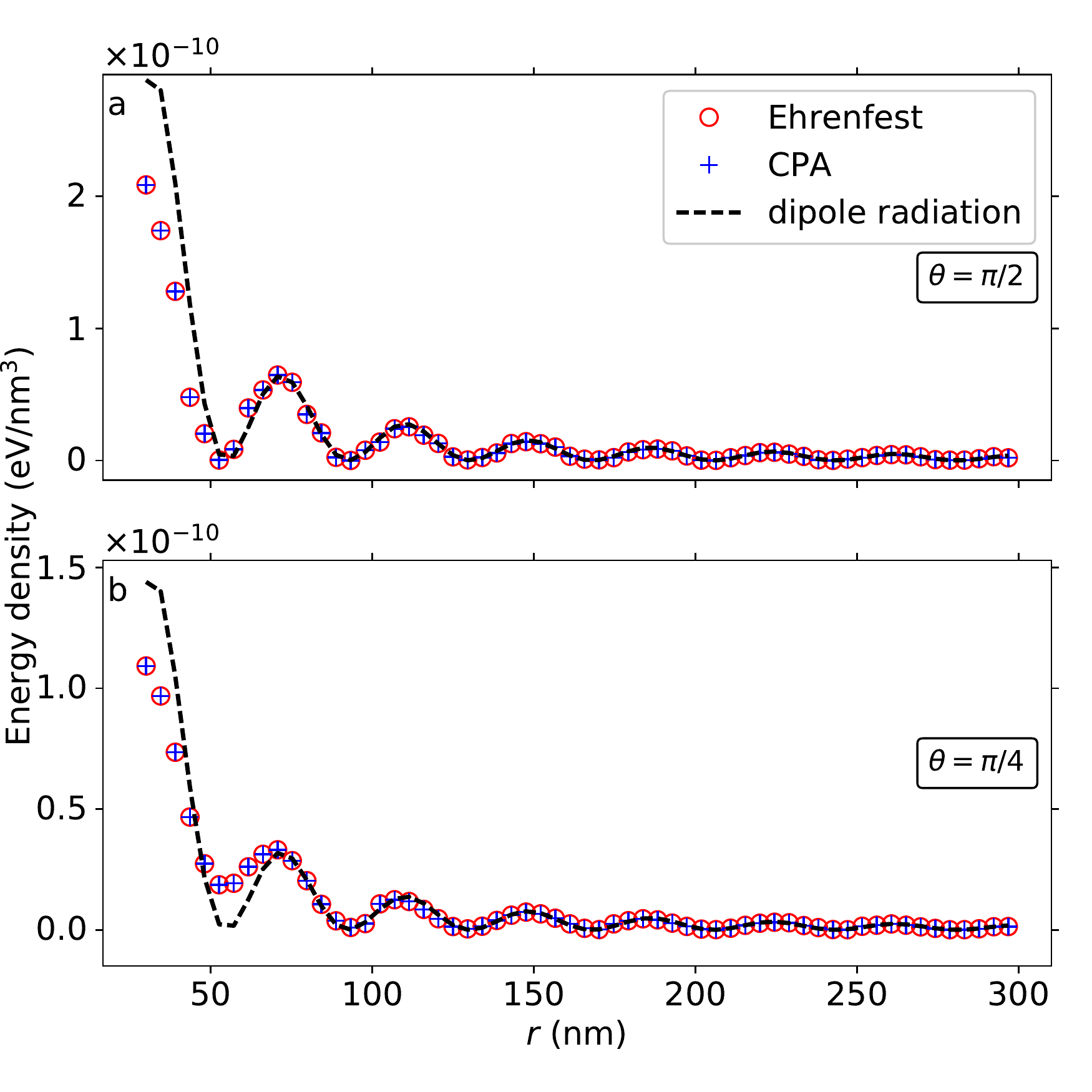}
	\caption{\label{fig:ehrenfest_3d_em_2} The energy density of the spontaneous EM field
(as predicted by Ehrenfest dynamics in 3D) versus  radius $r$ when $t=1.00$ fs.  The polar angle is (\textit{a}) $\theta = \pi/2$; (\textit{b}) $\theta = \pi/4$. All parameters are the same as in Fig. \ref{fig:ehrenfest_3d_em}. The radial distribution of EM energy density is the same for Ehrenfest and the CPA at short times and, just as in  Fig. \ref{fig:ehrenfest_3d_em}, these radial distributions agree with the classical dipole radiation result (provided the initial electronic state is $(C_1, C_2)$ = $(\sqrt{1/2}, \sqrt{1/2})$).}
	\end{figure}
	Finally, all of the Ehrenfest simulations above have been repeated in 3D. Overall, as shown in Fig. \ref{fig:ehrenfest_3d_2}, the results are qualitatively the same as in 1D.  However, as was emphasized in Sec. \ref{sec:spontaneous_decay}, the decay rate now depends cubically (and not linearly) on $\omega_0$. 

		Concerning the radiation of EM field in 3D, in Fig. \ref{fig:ehrenfest_3d_em}, we plot the energy density versus polar angle $\theta$ at $r=294$ nm when time $t = 1.00$ fs. For such a short time, Ehrenfest dynamics (red $\circ$) and CPA (blue $+$) agree exactly: both results depend on the polar angle $\theta$ through $\sin^2\theta$. These results are in very good agreement with  theoretical dipole radiation (black line, Eqn. \ref{eq:dipole_radiation_1}). Lastly, in  Fig. \ref{fig:ehrenfest_3d_em_2}, we plot the energy density as a function of the radial distance $r$ from the molecule, while keeping the polar angle fixed at $\theta=\pi/2$ (\textit{a}) and $\theta=\pi/4$ (\textit{b}).  Again, Ehrenfest dynamics (red $\circ$) and the CPA (blue $+$) agree with each other and give oscillating results that agree with Eqn. (\ref{eq:dipole_radiation_1}) for dipole radiation at asymptotically large distances ($r\gg\lambda\gg d$). 
Given that the Ehrenfest decay rate does not match spontaneous emission, one might be surprised
at the unexpected agreement between Ehrenfest and the CPA dynamics with the classical dipole radiation in Figs. \ref{fig:ehrenfest_3d_em}-\ref{fig:ehrenfest_3d_em_2}.  In fact, this agreement is somewhat coincidental (depending on initial conditions), as is proved in the Appendix.

	\section{Discussion}\label{sec:discussion}

	The results above suggest that, for their respective domains of applicability, both Ehrenfest dynamics and SQC can recover spontaneous emission. We will now test this assertion by investigating   the response to $(i)$ photo-induced dynamics and $(ii)$  dephasing.

	\subsection{An incoming pulse in one dimension }
	To address photo-induced dynamics, we imagine there is an incident pulse  at $t=0$ of the form:
	\begin{eqnarray}\label{eq:initial_pulse}
	\begin{split}
		\sqrt{\epsilon_0}E_z(x) &= -\frac{B_z(x) }{\sqrt{\mu_0}} \\
	&= A(b, k_0, x_0)e^{-b(x-x_0)^2}\cos(k_0x)
	\end{split}
	\end{eqnarray}
	Here,  $A(b,k_0,x_0)$ is an normalization coefficient with value
	\begin{equation*}
	A(b, k_0,x_0) = \sqrt{\frac{2U_0}{\sqrt{\pi/2b}(1+\cos(2k_0x_0)e^{-k_0^2/2b})}}
	\end{equation*}
	The total energy of incident pulse is  $U_0$. The parameter $b$ determines the width of the pulse in real space. $k_0$ defines the peak of the pulse in reciprocal space. $x_0$ represents the center of pulse at $t=0$.
	
	At time zero, the Fourier transform of $E_z(x)$  is:
	\begin{eqnarray}\label{eq:fourier_transform_E_y}
	\begin{split}
	E_z(k_x) &= \frac{1}{\sqrt{2\pi}}\int_{-\infty}^{\infty} dx\ E_z(x)e^{ik_xx} \\
	&= \frac{\epsilon_0 A(b, k_0, x_0)}{2\sqrt{2b}}\times  \\
	&\ \ \left( e^{-\frac{(k_x-k_0)^2}{4b}}e^{i(k_x-k_0)x_0} 
	+ e^{-\frac{(k_x+k_0)^2}{4b}}e^{i(k_x+k_0)x_0} \right )
	\end{split}
	\end{eqnarray}
	 $E_z(k_x)$ is the sum of two Gaussians centered at $k_x=\pm k_0$ with width $\sigma = \sqrt{2b}$. Qualitatively, if $b\ll k_0^2$,  $E_z(k_x)$ shows two peaks at $k_x=\pm k_0$; if $b \gg k_0^2$, $E_z(k_x)$ resembles a single large packet at $k_x=0$. For resonance  with the molecule,  $|E_z(k_x)|$ should be large at $\hbar c k_x = \hbar \omega_0$ (16.46 eV by default).

	\subsubsection{Electronic dynamics}
	
	\begin{figure}
		\includegraphics[width=8.5cm]{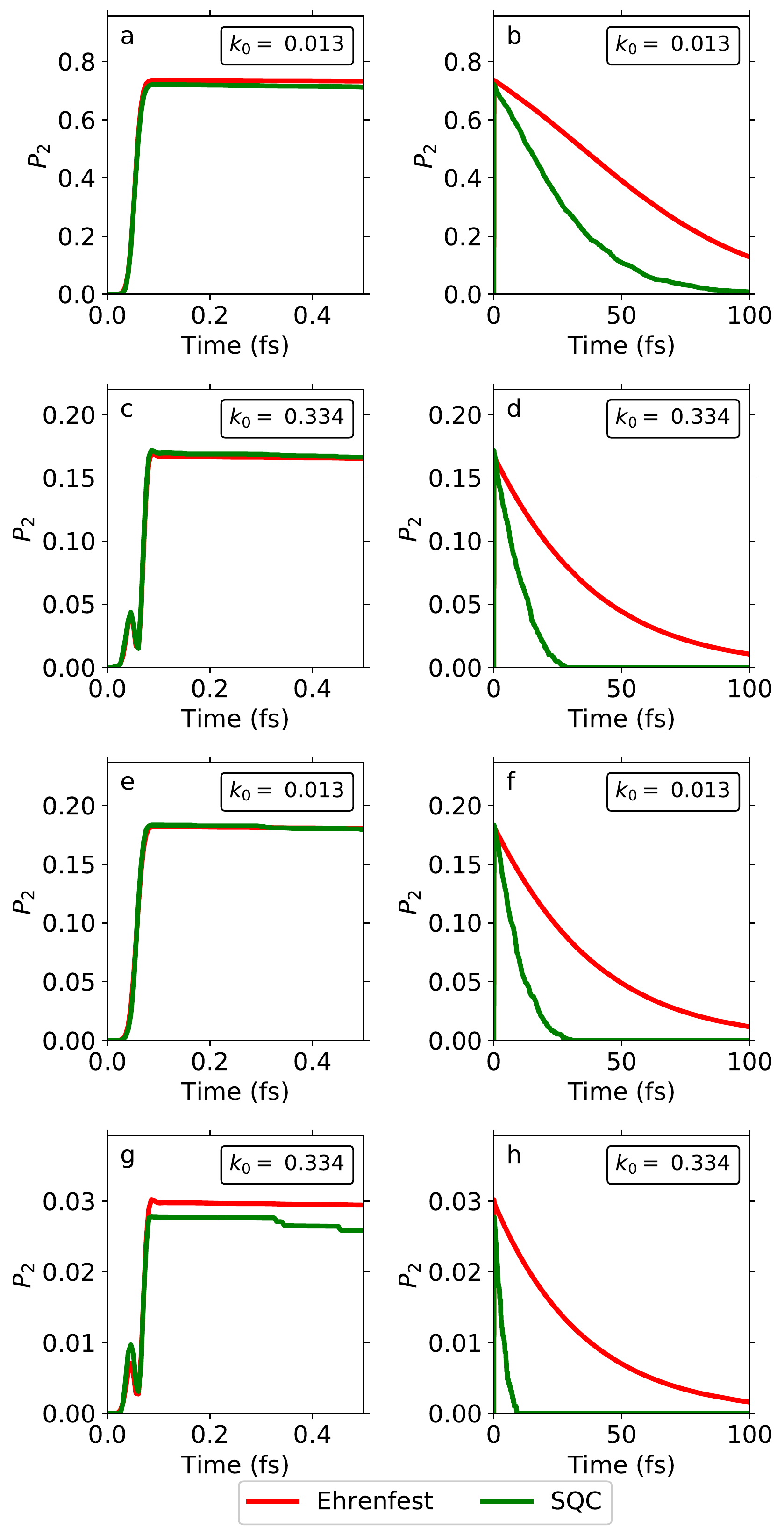}
		\caption{\label{fig:pulse_1d_U_4000} 
		A plot of the excited state electronic population $P_2$ as a function of time after exposure to an incident pulse of light. Early time dynamics are plotted on the \textit{left}, longer time dynamics is one the \textit{right}. Pulse parameters are listed in the table below.  
		Unreported parameters are set to their default values in Table \ref{tab:table1}. The initial electronic state $(C_1, C_2)$ = $(1, 0)$. Two methods are compared: Ehrenfest dynamics  (red line) and  SQC (green line). Note that SQC and  Ehrenfest dynamics disagree for long times, especially for weak pulses. See Fig. \ref{fig:pulse_1d_k_p2_relation}. For these simulations, we apply ABC's. Numerical results for Ehrenfest dynamics show that enforcing ABC's does not make any difference at all.}
   	 	\begin{tabular}{ccccc}
   	 	\hline 
   	 	No.& $U_0$ (keV) & $b (\text{\ nm}^{-1})$& $k_0 (\text{\ nm}^{-1})$ &$x_0$ (nm)\\ \hline 
   	 	(\textit{a-b})& 19.7 & 0.0556 & 0.013 & -15.0\\
   	 	(\textit{c-d})& 19.7 & 0.0556 & 0.334 & -15.0\\
   	 	(\textit{e-f})&  3.29 & 0.0556 & 0.013  & -15.0\\
   	 	(\textit{g-h})&  3.29 & 0.0556 & 0.334 & -15.0\\
   	 	\hline 
   	 \end{tabular}
	\end{figure}
	
		\begin{figure}
		\includegraphics[width=9cm]{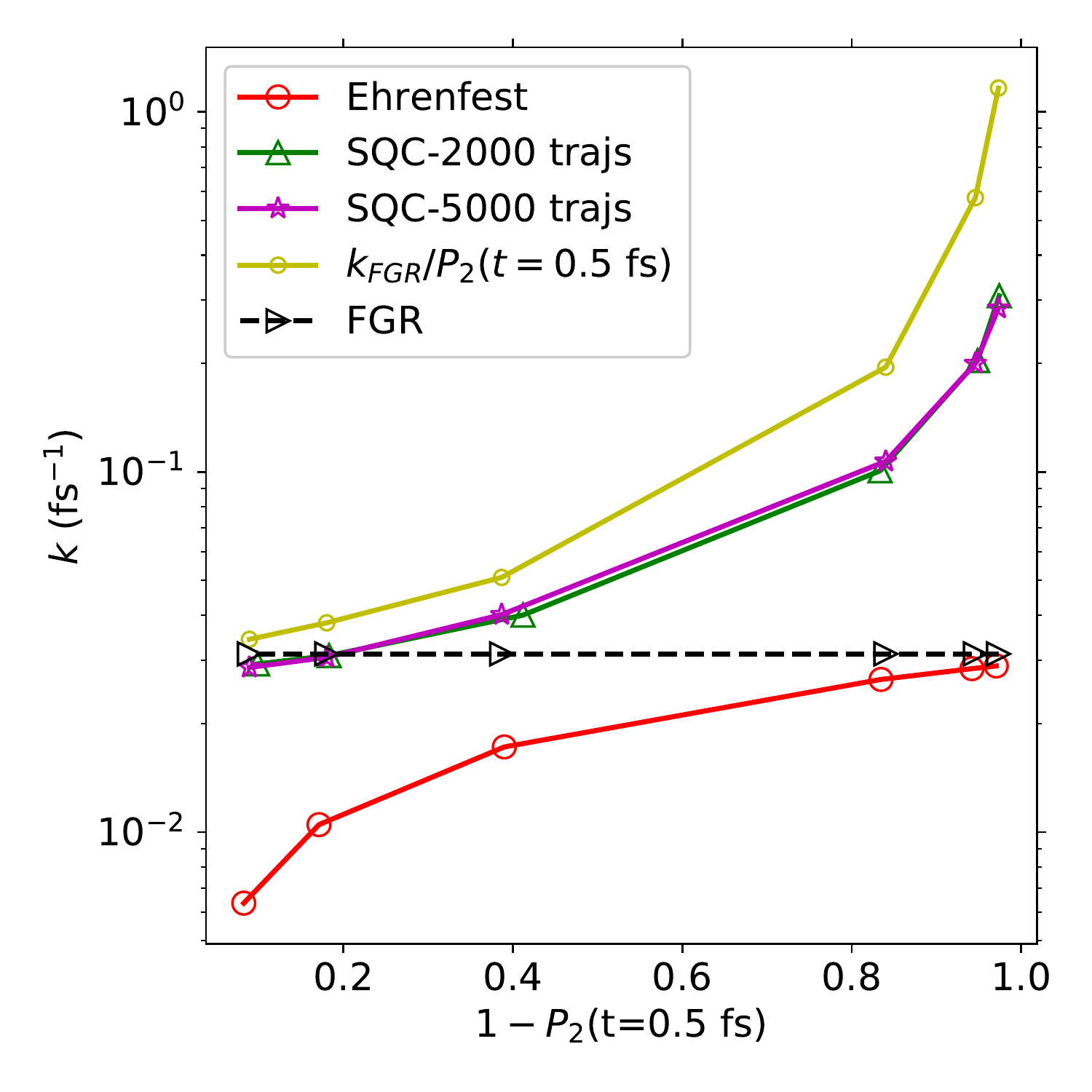}
		\caption{\label{fig:pulse_1d_k_p2_relation} The fitted decay rate $k$ versus $1-P_2(t=0.5 \text{ fs})$ following an incident pulse. Ehrenfest  rates
are basically identical with the spontaneous emission rates in Fig. \ref{fig:ehrenfest_1d_c1}. 
SQC yields the correct rate when the initial excited state population is close to one ( $P_2 \approx 1$), but strongly overestimates $k$ in the weak resonance regime ($P_2 \ll 1$). The behavior of SQC is roughly proportional to $k_{\text{FGR}}/P_2(t=0.5\text{ fs})$ (which goes to infinity as $P_2(t=0.5\text{ fs})$ goes to zero). Parameters for the incident pulse: $k_0 = 0.334\text{ nm}^{-1}$, $b = 0.0556\text{ nm}^{-2}$, $x_0 = -15.0\text{ nm}$ and $U_0$ varies from $3.29$ keV to $658$ keV. All other parameters are the same as in Fig. \ref{fig:pulse_1d_U_4000}. For these simulations, we apply ABC's. Numerical results for Ehrenfest dynamics show that enforcing ABC's does not make any difference at all. }
	\end{figure}

	In Fig. \ref{fig:pulse_1d_U_4000}, we plot the electronic population of the excited state as a function of time after exposure to incident pulses of different intensity ($U_0$) and wavevector ($k_0$); see  Eqn. (\ref{eq:initial_pulse}). We plot short and long times, on the left and right hand sides, respectively. For strong, resonant pulses, ($U_0 = 19.7$ keV, $k_0 = 0.013\text{ nm}^{-1}$), there is obviously a strong response (see \textit{a}-\textit{b}).  For strong, off-resonant pulses ($U_0 = 19.7$ keV, $k_0 = 0.334\text{ nm}^{-1}$), obviously the response is weaker. In both situations, SQC (green line) and Ehrenfest dynamics (red line)
	agree almost exactly for short times. At longer times, however, the SQC $P_2(t)$ value decays $\sim 2$ times faster than the Ehrenfest dynamics result. 

	Let us consider now weak pulses. In Fig. \ref{fig:pulse_1d_U_4000}\textit{e-h}, we plot the excited state population when the incident pulse  is weak ($U_0=3.29$ keV), keeping all other parameters unchanged.  Now, there is much less agreement between SQC and Ehrenfest dynamics, especially for long times.  Generally, SQC predicts a faster decay rate for $P_2(t)$ than Ehrenfest dynamics for small $|E_z(\omega_0/c)|$.
	
	The statement above is quantified in Fig. \ref{fig:pulse_1d_k_p2_relation}. Here, we vary $U_0$, which results in a change in the initial absorption (which is quantified by $1-P_2(t=0.5\text{ fs})$ on the $x$-axis).  This graph quantifies how the population decay on the excited state depends on the initial condition: the decay of $P_2$  decreases when the initial excited state population decreases. Obviously,  this Ehrenfest data is in complete agreement with Fig. \ref{fig:ehrenfest_1d_c1}. 

	Now, the new piece of data in Fig. \ref{fig:pulse_1d_k_p2_relation} is the SQC data. Here, we see that SQC behaves in a manner completely opposite to Ehrenfest: the decay of $P_2$  increases (sometimes dramatically) when the initial excited state population decreases.  Thus, for an initial state near $(1,0)$, the decay of $P_2$ is unphysically large according to SQC. At the same time, however, the decay of the state $(0,1)$ is very close to the FGR result (just as noted in Sec. \ref{sec:result}). Apparently, by including the zero point energy of the electronic state, SQC is able to include some aspects of true spontaneous decay, but the binning procedure introduces other unnatural consequences.  Future work on the proper binning procedure for SQC (triangles, squares, etc. \cite{Cotton2016}) must address this dilemma.

	\subsubsection{Distribution of the EM field}
	\begin{figure*}
		\includegraphics[width=18cm]{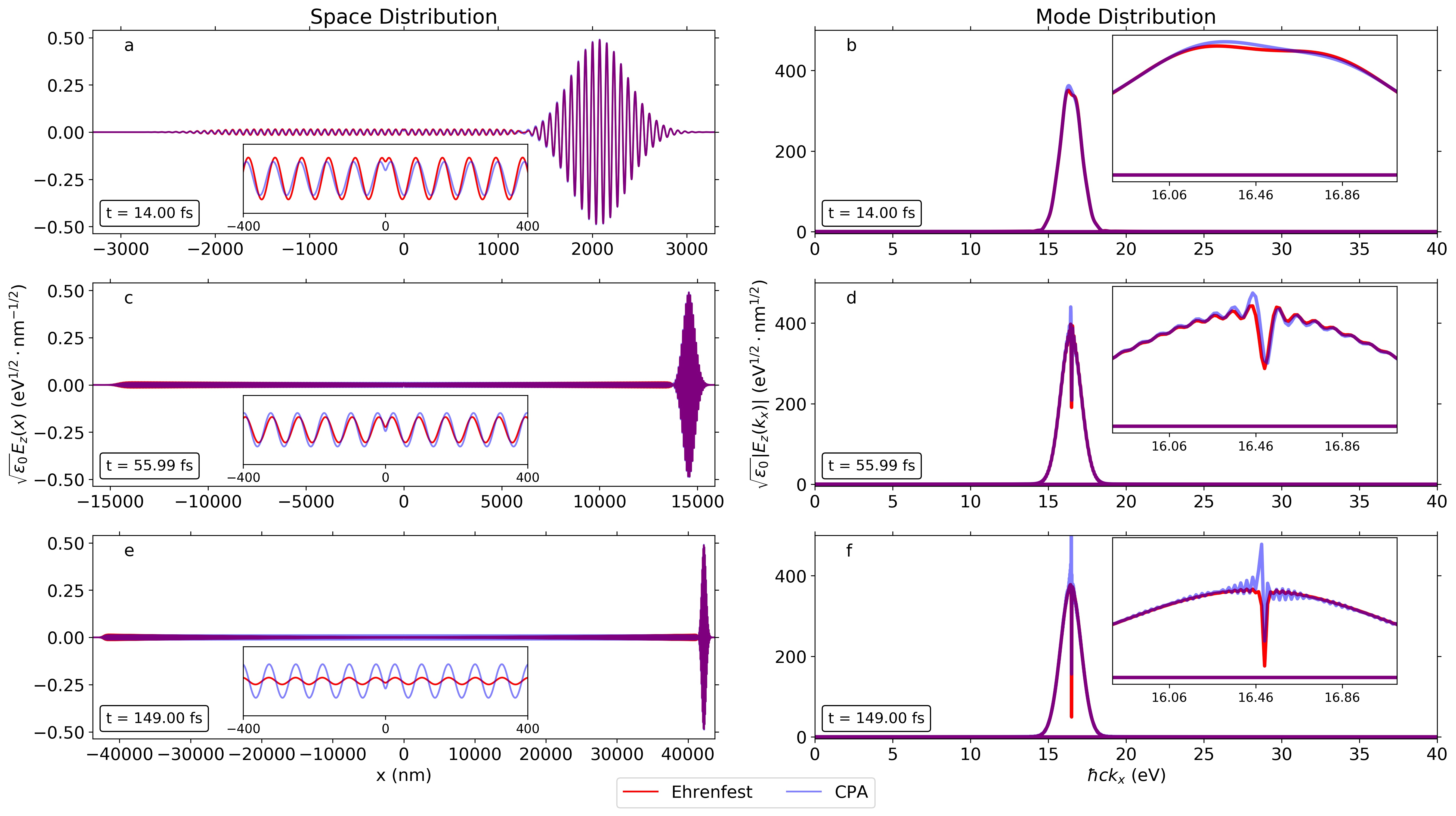}
		\caption{\label{fig:pulse_em_mode} For an incident pulse in 1D, we plot (\textit{left}) the spatial distribution of $E_z(x)$ at times (\textit{a}) 14.00 fs, (\textit{c}) 55.99 fs, and (\textit{e}) 149.00 fs; (\textit{right}) the mode distribution of $E_z$ in Fourier space at corresponding times. The inset figures zoom in on (\textit{left}) the ``molecule'' at the origin of the $x$-axis, and (\textit{right}) the two-level energy gap $\hbar \omega_0$ (here, 16.46 eV). Two methods are compared: Ehrenfest dynamics (red lines) and the CPA (light blue lines). Parameters for the incident pulse are $U_0=65.82 \text{eV}$, $k_0 = 0.08338\text{ nm}^{-1}$, $b = 5.56\times 10^{-6} \text{nm}^{-2}$ and $x_0 = -2098.6$ nm. All other parameters are the same as in Fig. \ref{fig:pulse_1d_U_4000}. Note that Ehrenfest and and CPA dynamics  agree at short times but disagree at long times when energy conservation becomes important.}
	\end{figure*}
	\begin{figure}
	\includegraphics[width=9cm]{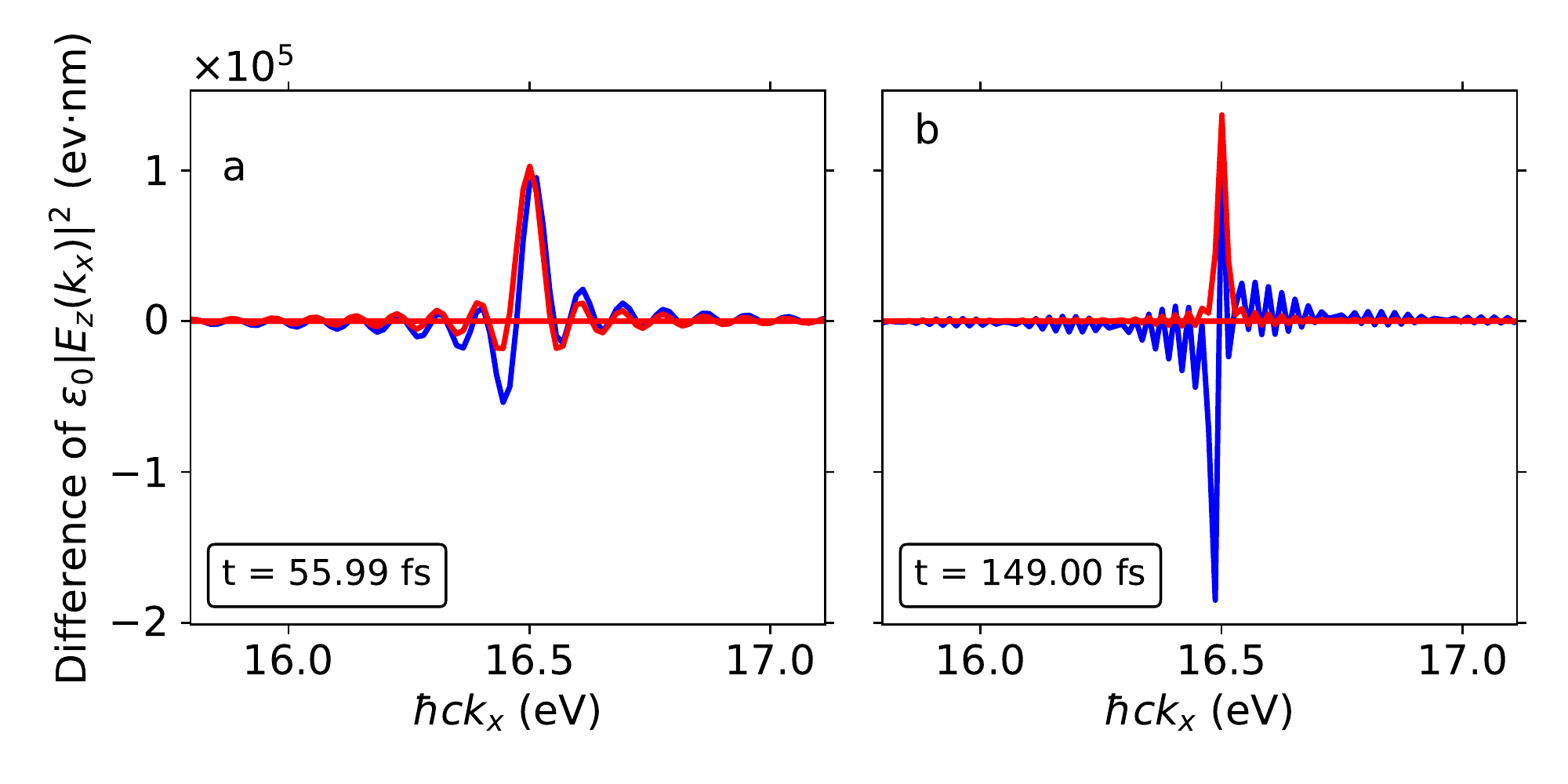}
	\caption{\label{fig:pulse_em_mode_difference} 1D Ehrenfest (red) and CPA (blue) absorption spectra at times (\textit{a}) 55.99 fs and (\textit{b}) 149.00 fs.  Spectra were obtained by subtracting $|E_z^{\text{free}}(k_x)|^2 - |E_z^{\text{Ehrenfest/CPA}}(k_x)|^2$.  Here,  $E_z^{\text{free}}$ denotes the freely propagated pulse (i.e. we set $\vJ$ to zero in Eqn. \ref{Maxwell}).  All simulation parameters are the same as in Fig. \ref{fig:pulse_em_mode}.  }
\end{figure}
	
	At this point, we should also comment on the EM field that is produced following incident radiation for the two-level system. Effectively, our results are consistent with Fig. \ref{fig:ehrenfest_final_em} above.  In Fig. \ref{fig:pulse_em_mode}, on the left, we plot  $E_z(x)$ versus $x$ in space at times 14.00 fs (\textit{a}), 55.99 fs (\textit{c}) and 149.00 fs (\textit{e}).   On the right hand side, we plot the Fourier transform $E_z(k_x)$ versus photon energy $\hbar ck_x$.  As above, we find that, for short times, Ehrenfest dynamics (red lines) and the CPA (light blue lines) are in good agreement. Thereafter, however, the agreement ends because only Ehrenfest dynamics obeys energy conservation.  At long times, Ehrenfest dynamics predicts an overall  dip (narrow decrease) in the electric field at the frequency of the two-level system (oscillator), while the CPA predicts an overall spike (narrow increase). Thus,  if we calculate the absorption spectrum of the molecule by subtracting the total transmitted signal from the freely propagated signal, as in Fig. \ref{fig:pulse_em_mode_difference}, only the Ehrenfest absorption spectrum is strictly positive; the CPA result makes no sense.	This state of affairs reminds us when and how we can use semiclassical theory for understanding light-matter interactions.
	
	Note that, for Fig. \ref{fig:pulse_em_mode_difference}, we are operating in the linear response regime: the incoming pulse energy $U_0$ is relatively weak. In Appendix C, we plot the absorption spectra for a few different incoming fields and demonstrate that the results are linear with $U_0$. We also show that standard linear response theory yields a good estimate of  the overall lineshape.

	\subsection{Dephasing effects} 
	In the present article, we have now shown that semiclassical theories -- Ehrenfest and SQC -- can both recover some elements of spontaneous emission, which is mostly thought to be a quantum effect\cite{Miller1978, Sukharev2011}. With this claim in mind, however, there is now one final subject that must be addressed, namely the role of dephasing. After all, in a large simulation with an environment, dephasing can and will occur; therefore one must wonder whether or not such dephasing will affect the rate of spontaneous emission.

	To answer this question, we have run several simple calculations that replace Eqn. (\ref{density_matrix_evolution}) by Eqn. (\ref{eq:1d_dump}),
	\begin{equation}\label{eq:1d_dump}
	\frac{d}{dt}\hrho(t) = -\frac{i}{\hbar}[\hH_s - \int d\vecr\  \vE(\vecr)\cdot \hP(\vecr),\  \hrho ] - 
	\begin{pmatrix}
	0 & \varsigma \rho_{12}\\ \varsigma^{\ast}\rho_{21} & 0 
	\end{pmatrix}
	\end{equation}
	Thus, we have propagated electron-photon dynamics by altering the electronic equation of motion but keeping the classical EM equations the same. $\varsigma$ in Eqn. (\ref{eq:1d_dump}) is an empirical dephasing rate: when $\varsigma = 0$, there is no dephasing and when $\varsigma > 0$ there is a finite rate of coherence loss between the two electronic states.  

\begin{figure}
		\includegraphics[width=9cm]{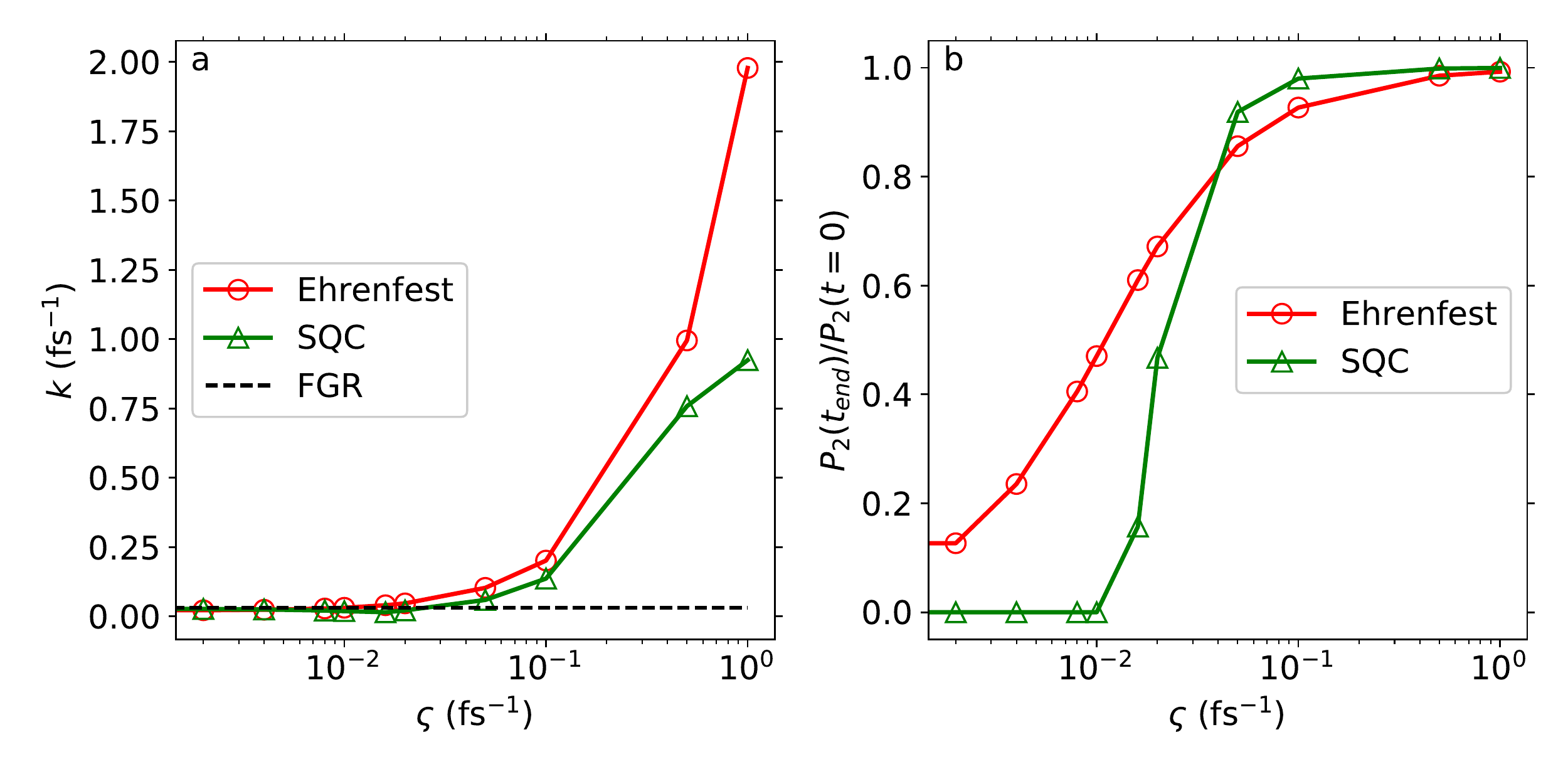}
		\caption{\label{fig:dump_1d} An analysis of the effects of dephasing on spontaneous emission
for Ehrenfest and SQC methods in 1D. A plot of  (\textit{a})the fitted decay rate $k$ as a function of the dephasing rate $\varsigma$; (\textit{b}) the normalized long time population of $P_2$, $P_2(t_{\text{end}})/P_2(t=0)$ as a function of $\varsigma$. 
All simulation parameters are set to their default values in Table \ref{tab:table1}; $t_{\text{end}}=400$ fs. The initial electric population for the excited state is set to  $P_2(0) = 1/2$ for Ehrenfest dynamics and $P_2(0) = 1$ for SQC dynamics. Note that both methods fail to recover spontaneous emission in the presence of strong dephasing. For these simulations, we apply ABC's. Numerical results for Ehrenfest dynamics show that enforcing ABC's does not make any difference at all. }
	\end{figure}

	In Fig. \ref{fig:dump_1d}\textit{a}, we plot the rate of spontaneous emission  $k$ as a function of the dephasing rate $\varsigma$.  When dephasing increases, the coherence between the electronic states is expected to decrease, and so the current should decrease, and thus the rate of spontaneous emission is expected to decrease as well. However, perhaps surprisingly, the fitted rate for establishing equilibrium also increases.

	Most importantly, in Fig. \ref{fig:dump_1d}\textit{b}, we plot the final population of the excited state $P_2(t_{\text{end}})$. As should be expected, the long term excited state population increases  (does not reduce to zero) when dephasing increases with either SQC or Ehrenfest dynamics.  This graph highlights the limitations of semiclassical methods: as currently implemented, one cannot include both spontaneous emission and dephasing.

	\section{Conclusion}\label{sec:conclusion}
	
	In this article, we have simulated the semiclassical dynamics of light coupled to a two-level electronic system with three different methods: Ehrenfest, the CPA and SQC.  Most results have been reported in one dimensional, but we have also considered Ehrenfest dynamics in 3D with absorbing boundary conditions. 
	As far as spontaneous emission is concerned, the CPA cannot  consistently recover the effect and violates the energy conservation. That being said,  Ehrenfest dynamics do predict spontaneous decay consistently, but only provided that we start in a non-trivial superposition state (with $C_1,C_2 \ne 0$).  Using electronic ZPE, SQC dynamics predicts spontaneous decay even with $C_1 = 0$. Both latter methods yield results fairly close to the correct FGR rate.  In all cases, unfortunately, spontaneous emission is destroyed when dephasing is introduced, which represents a fundamental limitation of semiclassical dynamics.

	Perhaps most interestingly, we have also studied photo-initiated excited dynamics and, in this case, we find very different dynamics as predicted by the different semiclassical methods. First, as far the EM field is concerned, we have demonstrated that Ehrenfest dynamics can recover the correct absorption spectra, at least qualitatively; at the same time, however,   CPA dynamics gives qualitatively incorrect spectra because the method ignores feedback and does not conserve energy. Second, and equally interesting, Ehrenfest dynamics predicts that the overall stimulated decay rate will depend smoothly on initial state $(C_1, C_2)$ but will approach the FGR rate in the weak resonance regime. Vice versa, SQC recovers FGR when $(C_1, C_2) = (0, 1)$ but overestimates the stimulated decay rate, sometimes by as much as a factor of 10 in the weak coupling limit. These SQC anomalies should be very important for designing improved binning protocols in the future\cite{Cotton2016}. At present, because the cost of SQC dynamics is roughly $1000$ times greater than Ehrenfest dynamics and because the method appears to fail for low intensity applied fields, further modification will likely be required before the method can be practical for large-scale simulations.

	Looking forward, many questions remain. $(i)$ There are {\bf many} other semiclassical methods for studying coupled nuclear electronic dynamics\cite{Subotnik2016, Habershon2013, Sun1997, Cao1994, Cotton2013}; will these methods give us new insight into electrodynamics? $(ii)$ Might we learn more about spontaneous emission by considering ZPE effects through RPMD-like algorithms\cite{PerezDeTudela2012}? $(iii)$ Will different semiclassical methods behave similarly or  differently with more than two electronic states?  $(iv)$  Can we converge multiple-spawning\cite{ben2000ab, martinez2006insights, tao2009ab, levine2009ab, kim2015ab} and/or MC-TDH\cite{beck2000multiconfiguration, meyer2006calculation, wang2003multilayer} calculations and generate exact quantum electrodynamical trajectories so that, in the future, we may benchmark other, less exact, semi-classical approximations?  And lastly, $(iv)$, are there other, new and non-intuitive features that will emerge when we study multiple pulses incoming upon a molecule? These questions will be answered in the future.
	\section*{Acknowledgments}
	This material is based upon work supported by the (U.S.)
	Air Force Office of Scientific Research (USAFOSR) PECASE
	award under AFOSR Grant No. FA9950-13-1-0157 (TL, HTC, JES),  AFOSR grant No. FA9550-15-1-0189 (MS), U.S. - Israel Binational Science Foundation Grant No. 2014113 (MS and AN) and the U.S. National Science Foundation Grant No. CHE1665291(AN). This work was also supported by the AMOS program within the Chemical Sciences, Geosciences and Biosciences Division of the Office of Basic Energy Sciences, Office of Science, US Department of Energy (TM).  The authors thank Phil Bucksbaum for very stimulating conversations.
	\section*{Appendix} 
\subsection{Connecting Ehrenfest Dynamics with Fermi's Golden Rule in 1D}\label{Appendix_A}
	We  now prove analytically that the spontaneous decay rate of Ehrenfest dynamics in 1D is exactly the  FGR result in the limit  that the initial excited state population is small ($P_2 \rightarrow 0$). 

	For Eqn. (\ref{Maxwell}), we can directly write down an analytic solution for $\vE(x)$ in one dimension using the well known solution for a wave equation with a source:
	\begin{eqnarray}\label{eq:analyticsolutionE}
	\begin{split}
		\vE(x, t) = &\frac{\omega_0}{c\epsilon_0} \{ \text{Im}\rho_{12}(0)\int_{x-ct}^{x+ct}dx'\vec{\xi}(x') \\
		&+ \int_{0}^{t}dt'\text{Im}\dot{\rho}_{12}(t')\int_{x-c(t-t')}^{x+c(t-t')}dx'\vec{\xi}(x')
		 \}
	\end{split}
	\end{eqnarray}
	Here, $\dot{\rho}_{12}$ is the time derivative of $\rho_{12}$.  If we average over many different initial electronic populations with different phases, $\text{Im}\rho_{12}(0) =0$,   the average coupling is simpler:
	\begin{eqnarray}\label{eq:H_12}
	\begin{split}
	H_{12}^{el} &= -\int dx \vE(x)\cdot\vec{\xi}(x) \\
	&= -\frac{\omega_0}{c\epsilon_0}\int_{-\infty}^{+\infty} dx \xi(x)\int_{0}^{t}dt'\text{Im}\dot{\rho}_{12}(t')\int_{x-c(t-t')}^{x+c(t-t')}dx'\xi(x')
	\end{split}
	\end{eqnarray}
	Here,  we have denoted $\xi(x) = |\vec{\xi}(x)|. $  Now, for simplicity, suppose the width of the molecule is infinitely small (i.e., a point-dipole approximation), $\xi(x) \approx \mu_{12}\delta(x)$. In such a case, Eqn. (\ref{eq:H_12}) can be simplified as:
	\begin{eqnarray}
	H_{12}^{el}  = -\frac{\omega_0}{c\epsilon_0}|\mu_{12}|^2 \text{Im}\rho_{12}(t),
	\end{eqnarray}
	and therefore, from Eqn. (\ref{density_matrix_evolution}),
	\begin{eqnarray}\label{eq:dP_2}
	\begin{split}
	\frac{dP_2}{dt} &= -\frac{dP_1}{dt} = \  \frac{2}{\hbar}H_{12}^{el}  \text{Im}\rho_{12}(t) \\ & =  -\frac{\omega_0}{c\epsilon_0\hbar}|\mu_{12}|^2\times 2\left[ \text{Im}\rho_{12}(t)\right]^2
	\end{split}
	\end{eqnarray}
	
	At this point, we make the weak coupling approximation, and assume that the off-diagonal terms in $H^{el}$ are infinitely small, so that $\rho_{12}(t)\approx\sqrt{P_1P_2}e^{-i\omega_0t}$ is a meaningful first order approximation. Eqn. (\ref{eq:dP_2}) then reads:
	\begin{eqnarray}\label{eq:dP_2_2}
	\begin{split}
	\frac{dP_2}{dt} & = -\frac{dP_1}{dt} = \  -2k_{\text{FGR}}  P_1P_2\sin^2(\omega_0t),
	\end{split}
	\end{eqnarray}
	where $k_{\text{FGR}} = \frac{\omega_0}{c\epsilon_0\hbar}|\mu_{12}|^2$ is the FGR spontaneous decay rate in 1D (see Eqn. \ref{eq:FGR_1d}).
	From Eqn. (\ref{eq:dP_2_2}), we can derive the instantaneous transfer rate plus an analytical solution for all times as follows.
	
	First, we consider the instantaneous behavior of Ehrenfest dynamics for $P_2$ within the time scale $\tau$ by integrating Eqn. (\ref{eq:dP_2_2}) over the time interval $[t, t+\tau]$,
	\begin{eqnarray}
	\ln\frac{P_2(t+\tau)}{P_2(t)} = -2k_{\text{FGR}}\int_{t}^{t+\tau}dt'P_1(t')\sin^2(\omega_0 t'),
	\end{eqnarray}
	where $2\pi/\omega_0\ll \tau \ll 1/k_{\text{FGR}}$. The time scale $\tau$ is taken to be much smaller the time scale of spontaneous decay ($\tau \ll 1/k_{\text{FGR}}$) so that $P_1(t')$ does not change much and $P_1(t')\approx P_1(t)$. Also, $\tau$ is much larger than the phase
	oscillating period ($2\pi/\omega_0\ll \tau$), therefore $\sin^2(\omega_0 t')$ can be viewed as a rapid oscillation and we approximate the
	integral by
	\begin{eqnarray}
	\int_{t}^{t+\tau}dt' \sin^2(\omega_0 t') = \left[ \frac{t'}{2} - \frac{\sin(2\omega_0t')}{4\omega_0}\right]_{t}^{t+\tau} \approx \frac{\tau}{2}
	\end{eqnarray}
	Then we have
	\begin{eqnarray}
	\ln\frac{P_2(t+\tau)}{P_2(t)} \approx -k_{\text{FGR}}P_1(t)\tau
	\end{eqnarray}
	As a result, we write Ehrenfest dynamics for $P_2$ in the form of an exponential decay
	\begin{eqnarray}
	P_2(t) = P_2(0)e^{-\kappa(t) t},
	\end{eqnarray}
	where the instantaneous decay rate is time-dependent
	\begin{eqnarray}
	\kappa(t) = k_{\text{FGR}}P_1(t)
	\end{eqnarray}
	On the one hand, for short times, the decay rate is proportional to the initial population $k_{\text{FGR}}P_1(0)$ as shown in Fig. \ref{fig:ehrenfest_1d_c1}b, and we can conclude that Ehrenfest dynamics recovers the FGR rate when $P_1(0)\rightarrow 1$.
	
	On the other hand, we may recast Eqn. (\ref{eq:dP_2_2}) in terms of
	the population difference, $\Delta P = P_2 - P_1$,
	\begin{eqnarray}
	\frac{d\Delta P}{dt} = -k_{FGR}(1-\Delta P^2 )\sin\omega_0t
	\end{eqnarray}
	Just as above, the instantaneous
	behavior within the time scale $\tau$ can be obtained by
	\begin{eqnarray}
	\left [\frac{1}{2}\ln(1+\Delta P) - \frac{1}{2}\ln(1 - \Delta P)\right ]_t^{t+\tau} \approx -k_{\text{FGR}}\frac{\tau}{2}
	\end{eqnarray}
	Hence, we find an analytical form for $P_2$ according to Ehrenfest dynamics:
	\begin{eqnarray}\label{eq:ehrenfest_analytical}
	P_2(t) = \frac{e^{-k_{\text{FGR}}t}}{\frac{P_1(0)}{P_2(0)} + e^{-k_{\text{FGR}} t}}
	\end{eqnarray}
	For short times, we take $t\rightarrow 0$ and find that the instantaneous decay rate is also proportional to $P_1(0)$. For the initial population $P_1(0)=P_2(0)=1/2$, as was
	considered in Fig. \ref{fig:ehrenfest_1d_1}, the analytical solution becomes $P_2(t) = e^{-k_{\text{FGR}}t}/(1 + e^{-k_{\text{FGR}}t})$. This formula agrees with the numerical result in Fig. \ref{fig:ehrenfest_1d_1}.
	



\subsection{Connecting Ehrenfest Dynamics with classical dipole radiation in 3D}\label{Appendix_B}

		Here, we show that Ehrenfest dynamics agrees with classical dipole radiation at short times assuming that the initial conditions satisfy $(|C_1|, |C_2|)$ = $(\sqrt{1/2}, \sqrt{1/2})$. First, consider classical dipole radiation, and let the 
oscillating dipole (in the $z$-direction) be situated at the origin.  The current takes the form $\vec{I} = -q \omega\sin(\omega t + \phi)\hat{e}_z$  and if the dipole width $d$ is small enough, the current density  is
		\begin{equation}\label{eq:J_dipole_radiation}
		 \vJ(\vecr) = \lim_{d\rightarrow 0} \left[ d\cdot \vec{I}\right]\delta(\vecr) = -\mu_{12} \omega\sin(\omega t+ \phi)\delta(\vecr)\hat{e}_z
		\end{equation}
This is the source that acts as input for Maxwell's equations and yield classical dipole radiation.

		 Second, consider Ehrenfest dynamics. Now, $\vJ(\vecr)$ takes the form in Eqn. (\ref{eq:J_simplified}). If we take the weak coupling approximation, i.e. we assume  that $\rho_{12} \approx \sqrt{P_1P_2}e^{i\omega_0 t}e^{i\phi}$, and we further make the point dipole approximation, $\xi(\vecr) \approx \mu_{12}\delta(\vecr)$, then   Eqn. (\ref{eq:J_simplified}) becomes
		 \begin{eqnarray}\label{eq:J_ehrenfest_approximations}
		 \vJ(\vecr) = -2\sqrt{P_1P_2} \omega_0 \mu_{12} \sin(\omega_0 t+\phi)\delta(\vecr)\hat{e}_z
		 \end{eqnarray}
		 Lastly, if the initial electronic state satisfies $(|C_1|, |C_2|)$ = $(\sqrt{1/2}, \sqrt{1/2})$, then  $P_1P_2 = 1/4$. Thus, this initial electronic state guarantees that Eqns. (\ref{eq:J_dipole_radiation}) and (\ref{eq:J_ehrenfest_approximations}) will be identical at short times: the EM field from Ehrenfest dynamics will agree with classical dipole radiation exactly.  This exact agreement will fail for other initial states or at long times. Even though both methods have the same geometric form, in general, Ehrenfest dynamics would need to be rescaled to match classical dipole radiation in absolute value. 
\subsection{Absorption spectra with different incoming field intensities}
	\begin{figure*}
	\includegraphics[width=18cm]{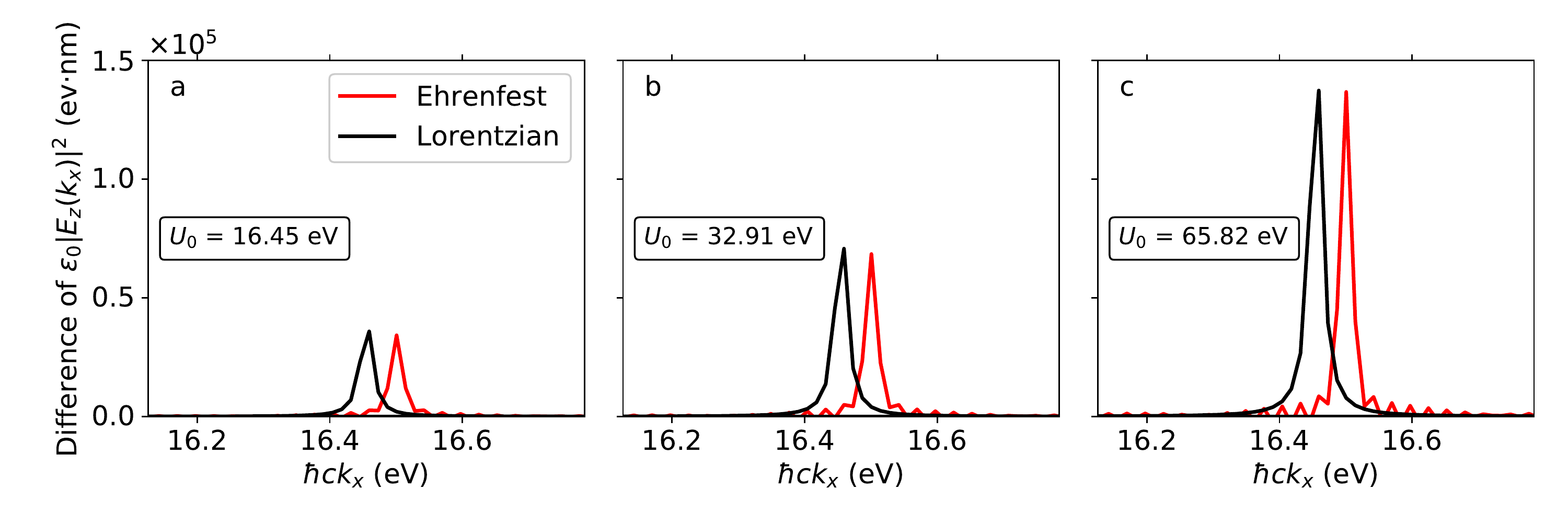}
	\caption{\label{fig:pulse_compare_abc} 1D simulated Ehrenfest (red)  absorption spectra for different incoming fields at time 149.00 fs.   Spectra were obtained in the same manner as in Fig. \ref{fig:pulse_em_mode_difference} while varying the incoming energy ($U_0$) of the incident pulse.  The value of $U_0$ is chosen to be: ($a$)  16.45 eV,  ($b$)  32.91 eV and  ($c$)  65.82 eV.  Note that the overall signal is linearly proportional to $U_0$
		and the lineshape width is nearly a Lorentzian centered at $\omega_0$ with width equal to the Fermi golden rule rate (black).
	 }
\end{figure*}
In this subsection, we plot the absorption lineshape for a variety of different incoming fields and prove that  the data in Fig. \ref{fig:pulse_em_mode_difference} is occurring in the linear regime.  Indeed, according to Fig. \ref{fig:pulse_compare_abc}, the overall absorption signal is linearly proportional to the incoming energy $U_0$. The absorption lineshape can be recovered approximately by simply assuming a Lorentzian signal with width $k_{\text{FGR}}$ and a uniform fitting for the total norm. Note that there is a small shift in the maximal signal location: according to Ehrenfest dynamics, the peak is centered at $\sqrt{\omega_0^2 + \Delta^2}$ (rather than $\omega_0$) where $\Delta$ is the time-averaged off-diagonal coupling in the Hamiltonian $\hH^{el}$. See Eqn. (\ref{eq:semi-H}).
	
%
	
\end{document}